\documentclass[fleqn,usenatbib]{mnras}

\usepackage{newtxtext,newtxmath}

\usepackage[T1]{fontenc}
\usepackage{orcidlink}

\DeclareRobustCommand{\VAN}[3]{#2}
\let\VANthebibliography\thebibliography
\def\thebibliography{\DeclareRobustCommand{\VAN}[3]{##3}\VANthebibliography}

\usepackage{graphicx}	
\usepackage{amsmath}	

\usepackage{color,soul}  

\include{Figs_hSN}

\newcommand{\kms}{\,km\,s$^{-1}$}	
\newcommand{\CIV}{\ion{C}{iv}}
\newcommand{\HeII}{\ion{He}{ii}}
\newcommand{\CIII}{\ion{C}{iii}]}
\newcommand{\SiIII}{\ion{Si}{iii}]}
\newcommand{\AlIII}{\ion{Al}{iii}}
\newcommand{\MgII}{\ion{Mg}{ii}}
\newcommand{\aro}{$\alpha_{\text{ro}}$} 
\newcommand{\logR}{$\log_{10} R$}  
\newcommand{\Lbol}{$L_{\text{bol}}$}  

\newcommand{\figheiilum}{\ref{fig:heii_3panel}}
\newcommand{\figheiipd}{\ref{fig:heii_3panel}}
\newcommand{\figtracks}{\ref{fig:heii_3panel}}

\renewcommand{\hl}[1]{#1}

\title[LOFAR-detected quasars in \CIV\ emission space]{Placing LOFAR-detected quasars in \CIV\ emission space: implications for winds, jets and star formation}

\author[Rankine et al.]
{Amy L. Rankine$^{1}$\thanks{E-mail: alrankine@ast.cam.ac.uk}\orcidlink{0000-0002-2091-1966}, 
James H. Matthews$^{1}$\orcidlink{0000-0002-3493-7737}, 
Paul C. Hewett$^{1}$\orcidlink{0000-0002-6528-1937}, 
Manda Banerji$^{2,1,3}$\orcidlink{0000-0002-0639-5141} 
\newauthor Leah K. Morabito$^{4,5}$\orcidlink{0000-0003-0487-6651} 
and Gordon T. Richards$^{6}$\orcidlink{0000-0002-1061-1804} \\
$^1$Institute of Astronomy, University of Cambridge, Madingley Road, Cambridge, CB3 0HA\\
$^{2}$School of Physics \& Astronomy, University of Southampton, Southampton, SO17 1BJ, UK\\
$^3$Kavli Institute for Cosmology, University of Cambridge, Madingley Road, Cambridge CB3 0HA, UK\\
$^4$Centre for Extragalactic Astronomy, Department of Physics, Durham University, Durham, DH1 3LE, UK\\
$^5$Institute for Computational Cosmology, Department of Physics, University of Durham, South Road, Durham DH1 3LE, UK \\
$^6$Department of Physics, Drexel University, 32 S. 32nd Street, Philadelphia, PA 19104, USA
}

\date{\today}

\pubyear{2021}

\begin{document}
\label{firstpage}
\pagerange{\pageref{firstpage}--\pageref{lastpage}}
\maketitle

\begin{abstract}
We present an investigation of the low-frequency radio and ultraviolet properties of a sample of $\simeq$10\,500 quasars from the Sloan Digital Sky Survey Data Release 14, observed as part of the first data release of the Low-Frequency-Array Two-metre Sky Survey. The quasars have redshifts $1.5 < z < 3.5$ and luminosities $44.6 < \log_{10}\left(L_{\text{bol}}/\text{erg\,s}^{-1}\right) < 47.2$. We employ ultraviolet spectral reconstructions based on an independent component analysis to parametrize the \CIV\,$\lambda$1549-emission line that is used to infer the strength of accretion disc winds, and the \HeII\,$\lambda$1640 line, an indicator of the soft X-ray flux. We find that \hl{radio-detected quasars are found in the same region of {\ion{C}{iv}} blueshift versus equivalent-width space as radio-undetected quasars}, but that the loudest, most luminous and largest radio sources exist preferentially at low \CIV\ blueshifts. Additionally, the radio-detection fraction increases with blueshift whereas the radio-loud fraction decreases. In the radio-quiet population, we observe a range of \HeII\ equivalent widths as well as a Baldwin effect with bolometric luminosity, whilst the radio-loud population has mostly strong \HeII, consistent with a stronger soft X-ray flux. The presence of strong \HeII\ is a necessary but not sufficient condition to detect radio-loud emission suggesting some degree of stochasticity in jet formation. Using energetic arguments and Monte Carlo simulations, we explore the plausibility of winds, compact jets, and star formation as sources of the radio quiet emission, ruling out none. The existence of quasars with similar ultraviolet properties but differing radio properties suggests, perhaps, that the radio and ultraviolet emission is tracing activity occurring on different time-scales.
\end{abstract}




\section{Introduction}

Quasars and active galactic nuclei (AGN) are widely considered important in galaxy formation models, interacting with their host galaxies via collimated radio jets or wide-angled winds launched from the accretion disc. Both winds and jets are produced across all black hole masses; however, their connection to accretion processes and disc physics is unclear.

Accretion disc winds are most evident via the broad absorption lines (BALs) observed in $\simeq$20 per cent of quasar ultraviolet/optical spectra \citep{Hewett03}. The intrinsic, extinction-corrected fraction of quasars that are classed as BAL quasars is expected to be $\simeq$40 per cent \citep{Dai20082MASSBALQSOs, Allen2011AFraction}, with the majority exhibiting absorption of the highly ionized species \CIV\,$\lambda$1549. Also evidence of disc winds is the blueshifting of the \CIV\ emission line \citep[e.g.,][]{Gaskell1982AMotions, Sulentic2000EigenvectorNuclei, Leighly2004,richards_broad_2002, richards_unification_2011}. It is not yet clear how direct the connection is between the \CIV\ blueshift and the BALs; however, \citet{Rankine2020} reported strong correlations between the BAL and emission-line properties and suggested that BAL quasars are drawn from the same parent population as non-BAL quasars.

AGN jets are often identified and studied through their radio emission, and radio sources can be categorized based on their morphology, radio-loudness, and optical spectra (among other properties). \hl{Results showing bimodality in the distribution of radio-loudness} -- the ratio of radio to optical luminosity -- by \citet{kellermann_vla_1989} lead to the authors defining the radio-loud population of quasars \citep[see also][]{Strittmatter1980}. A series of studies have investigated the distribution of radio-loudness and whether or not a clear bimodality exists. Generally, it is clear that the distribution cannot be fit by a single Gaussian, but the evidence for a true bimodality or dichotomy is debated \citep{Falcke1996,Brotherton2001,ivezic_optical_2002,White2007,Miller2011,Balokovic2012,Gurkan2019}.
What is known is that the radio-loud population is dominated by powerful jets; however, the importance of compact jets, winds, and star formation in producing the radio emission of the radio-quiet quasars is still unclear \citep[see][for a review]{panessa_origin_2019}.

The radio-loud AGNs can be split into high and low excitation radio galaxies (HERGs and LERGs) as determined by the optical emission line properties \citep{laing_spectrophotometry_1994,tadhunter_nature_1998,Buttiglione10,best_fundamental_2012}. HERGs typically refer to AGNs in `quasar-mode' in which the black hole is thought to accrete from an optically thick and geometrically thin accretion disc. This mode can also be called `radiative-mode' owing to the disc's radiative efficiency. LERGs, on the other hand, are observed to have highly energetic radio jets but an absence of strong emission lines that are otherwise present in HERGs. \citet{best_fundamental_2012} find HERGs and LERGs across all radio luminosities, although with HERGs found preferentially at high radio luminosities, and they suggest that Eddington fraction may be the main driver of the HERG/LERG dichotomy. Additionally, the \citet{Fanaroff1974} classification scheme splits radio galaxies into Fanaroff–Riley type I (FR I) or Fanaroff–Riley type II (FR II) based on their morphology. \hl{The morphological class depends on radio luminosity, with FR II sources more common at high luminosity} \citep{Fanaroff1974,LedlowOwen1996}, \hl{implying that jet power is one important factor, although more recent results have shown that a luminosity break between FR I/II sources is far from clear} \citep{Best2009, Mingo2019}. It has been suggested that the FR I/II dichotomy might be driven by accretion rate or black hole spin \citep{Baum1995}, or differences in the host galaxy or environment \citep{Bicknell1995, Kaiser2007}. There is growing evidence that environment is important \citep{Hill1991, Gendre2013, Miraghaei2017, Mingo2019}, whereby jets with similar radio powers are more likely to be disrupted in denser environments thus becoming morphologically classed as FR Is \citep[see e.g.][for discussions]{Mingo2019,hardcastle_radio_2020}. 

The radio-quiet sources are arguably less well understood than the radio-loud ones, since they are fainter and harder to spatially resolve. Star formation produces radio emission \citep[e.g][]{condon_radio_1992,thompson_magnetic_2006,becker_cosmic_2009}, and jets and star formation must contribute at some level to the radio-quiet emission, with possible other contributions from disc winds \citep{stocke_radio_1992,blundell_origin_2007,zakamska_quasar_2014} and X-ray coronae \citep{laor_origin_2008,raginski_agn_2016}. A review of these processes is provided by \cite{panessa_origin_2019}. Whatever the origin of the emission, properly characterising the radio-quiet population is critical for understanding the connection between quasar outflows and star formation. In addition, the well-known correlation between far-infrared luminosity and radio luminosity, the far-infrared radio correlation \citep{helou_thermal_1985,yun_radio_2001,calistro_rivera_lofar_2017,gurkan_lofar/h-atlas:_2018,read_far-infrared_2018} means that radio luminosity can be used as a star formation rate estimator in certain cases; understanding the level at which this correlation is contaminated from AGN-driven mechanisms is again important. 

Although there have been attempts to combine winds and jets within a unified framework \citep[e.g.][]{Giustini2019}, the issue of how, in detail, jets and winds are produced, related, and connected to the accretion disc remains fundamentally unclear. The \CIV\ blueshift can be used as an indicator of wind strength, with quasars exhibiting large and positive blueshifts thought to host stronger accretion disc winds. Observations of radio-loud quasars act as a probe of jet physics, whereas the \CIV\ emission space is used to infer properties of the accretion disc, broad-line region (BLR) and disc winds. Combining these data therefore allows the connection between the AGN disc, jets, and winds to be studied. \hl{By investigating the radio properties of Sloan Digital Sky Survey (SDSS) DR7 quasars from the Faint Images of the Radio Sky at Twenty-Centimetres (FIRST) survey} \citep{Becker1995}, \citet{richards_unification_2011} \hl{were able to show that radio-loud quasars are concentrated at low} \CIV\ \hl{blueshifts, thus suggesting radio-loud quasars often have little to no outflowing wind.} This behaviour is broadly consistent with the relative scarcity of radio-loud BAL quasars \citep{stocke_radio_1992,becker_properties_2000,White2007,Morabito2019}.
\citet{richards_unification_2011} also found radio-quiet quasars with low blueshifts and, in fact, very similar \hl{UV spectral energy distributions (SEDs)} to the radio-loud quasars. \citet{kratzer_mean_2015}, also using FIRST and SDSS, were able to show that the mean radio-loudness decreases with increasing \CIV\ blueshift and argue that radio-loud and radio-quiet quasars should not be compared without first taking into account non-radio properties. \citet{Stone2019} investigated instead the narrow \CIV\ absorbers associated with the quasar and found them to be as common in radio-quiet quasars as they are in radio-loud quasars, suggesting that whatever physics governs the associated absorbers -- whether they are evidence of a failed accretion disc wind \citep[e.g.,][]{Vestergaard2003} or winds from star formation on kilo-parsec scales \citep{Barthel2017} -- it is unrelated, at least directly, to the radio emission \citep[see also][who found no correlation between the number of absorbers in quasar spectra and radio-loudness]{Chen2020}. 

Owing to the sensitivity of FIRST, \citet{richards_unification_2011} were only able to investigate the relationship between the UV and radio properties of the radio-loud population. With the advent of the high-sensitivity Low-Frequency Array \citep[LOFAR;][]{van_haarlem_lofar_2013}, it is now possible to probe the radio-quiet population in more detail \citep[e.g.][]{gurkan_herschel-atlas_2015,gurkan_lofar/h-atlas:_2018,Gurkan2019,Morabito2019,Rosario2020}. By combining LOFAR data with the \CIV\ emission line, we can therefore consider the relative contributions of winds, star formation, and compact jets in the radio-quiet population and their relation to disc and outflow processes. In addition, we can confirm whether or not the radio-loud population behaves in the same way when observed at lower radio frequencies. Such an approach can, in principle, lead to greater insight into quasar feedback and fuelling by untangling the radio emission produced by star formation from that of jets and winds, while simultaneously studying the connection between star formation and the quasar accretion properties. 

In this study, we have investigated the relationship between the low-frequency radio emission and the rest-frame UV properties, in particular the \CIV\ and \HeII\ emission line properties of quasars observed by SDSS and part of the \hl{LOFAR Two-metre Sky Survey (LoTSS)}. The paper is structured as follows. In Section~\ref{sec:data}, we describe the observational data used in this study and the \hl{independent component analysis (ICA)-based} spectrum-reconstruction scheme employed to parametrize the UV emission lines. The results are presented in Section~\ref{sec:results} before a discussion of their implications in Section~\ref{sec:discussion}. We employ vacuum wavelengths throughout the paper and Lambda cold dark matter cosmology with $h_0=0.71$, $\Omega_\text{M}=0.27$, and $\Omega_\Lambda=0.73$. We define the spectral index, $\alpha$, such that flux density $F_\nu \propto \nu^{\alpha}$, and use this definition throughout the paper.

\begin{figure*}
    \centering
    \includegraphics[width=\linewidth]{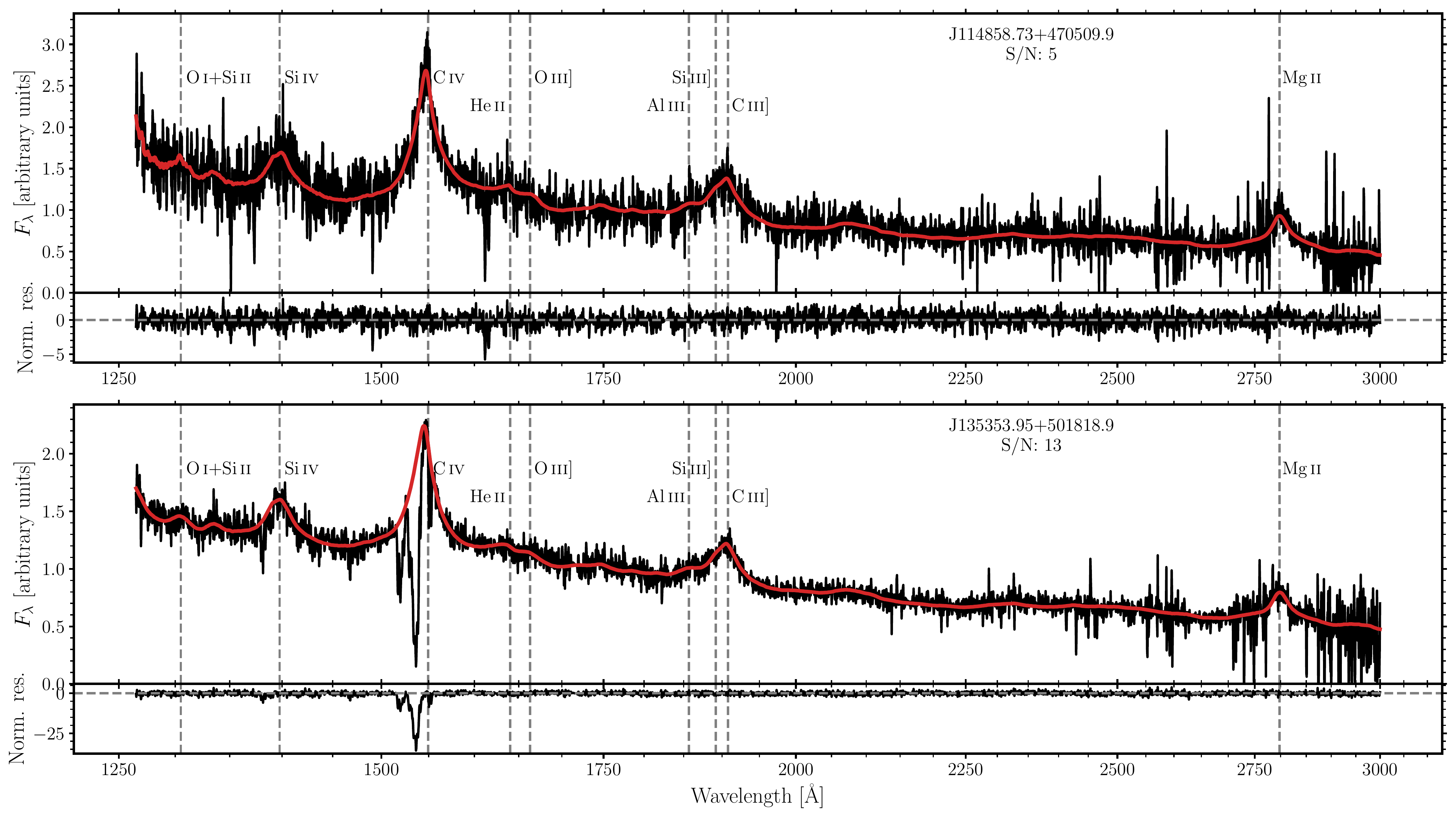}
    \caption{Two examples of SDSS DR14 spectra and the reconstructions from \citet{Rankine2020}. The top panels contain an example of a low-S/N (S/N $\simeq5$) quasar spectrum (black) and its reconstruction overplotted in red. The residuals normalized by the noise [i.e., (spectrum$-$reconstruction)/noise] are also shown. In low-S/N spectra, the reconstructions provide more robust measurements of the emission line parameters, in particular the \CIV\ blueshift (calculated from the flux-weighted median wavelength of the \CIV\ emission) and equivalent width (see Section~\ref{sec:CIVres} for details). The bottom panels contain an example of a quasar spectrum with significant absorption of the \CIV\ line. In such cases, the ICA components use priors based on the properties of the \CIII\,$\lambda1909$ complex to reconstruct the intrinsic \CIV\ emission.}
    \label{fig:spec}
\end{figure*}

\section{Observational data}
\label{sec:data}
The LoTSS \citep{Shimwell2017} is a low-frequency (144\,MHz), high-sensitivity (71\,$\mu$Jy\,beam$^{-1}$ median) survey with the aim of observing the entire northern sky. The first data release \citep[LDR1;][]{Shimwell2019} covers over 400\,deg$^2$ of the \hl{Hobby-Eberly Telescope Dark Energy Experiment (HETDEX) Spring field} and contains almost 320\,000 radio sources. The value-added catalogue contains optical identifications and morphological classifications \citep{Williams2019} as well as photometric redshifts and rest-frame magnitudes by combining Pan-STARRS and WISE data \citep{Duncan2019}. Using data from LoTSS allows us to investigate the low-frequency radio properties of SDSS quasars and examine trends with their emission line properties. One of the great advantages of LOFAR is its lower observing frequency compared to surveys such as the FIRST survey \citep{Becker1995}. For 5$\sigma$ detections, FIRST has a flux limit of 1\,mJy at 1.4\,GHz. LOFAR meanwhile, has a flux limit of 0.35\,mJy at 144\,MHz \citep{Rosario2020}. However, LOFAR is effectively 10 times more sensitive than FIRST to a compact radio source with $\alpha=-0.7$ \citep{Shimwell2019}, with the difference greater for steeper spectrum sources. 

Our quasar sample was defined by identifying quasars in the SDSS DR14Q catalogue \citep{Paris2018} lying within the $\simeq$400\,deg$^2$ area of LDR1. To do this, we used the multi-order coverage map generated by \cite{Morabito2019}, allowing us to match the LoTSS DR1 sample with SDSS DR14 quasars \hl{using the Pan-STARRS positions in the value-added catalogue from} \citet{Williams2019}. \hl{SDSS DR16Q} \citep{Lyke2020} \hl{was published during the time of writing. Our analysis for the DR14 quasars is made possible by our previous work on reconstructing the spectra }\citep{Rankine2020}\hl{. Additionally, using DR16 would lead to an increase in total sample size of 15 per cent and an increase in the detected sample of only 8 per cent within the redshift range of interest. Thus, using DR16 would require a substantial additional effort without changing our overall results. For these reasons, we chose not to use DR16.}
The spectrum signal-to-noise ratio (S/N) of the SDSS DR14Q quasars covers an extended range, with bright objects observed with S/N>20 per 69\kms\ pixel while quasars detected close to the SDSS-survey magnitude-limit possess S/N $\lesssim2$. 

The observed \CIV\,$\lambda$1549-emission properties depend both on the accuracy of the quasar systemic redshift estimates and the spectrum S/N in the vicinity of the \CIV-emission. To improve the accuracy of the \CIV-emission properties the same approach as described in \citet{Rankine2020} was adopted. Systemic redshifts were obtained using ICA reconstructions of the \MgII\,$\lambda$2800 emission and the emission complex of \CIII\,$\lambda$1908, \SiIII\,$\lambda$1892, \AlIII\,$\lambda$1857 \citep[and \ion{Fe}{iii} UV34;][]{Temple2020}. The scheme deliberately avoids utilising the \CIV\,$\lambda$1549 emission with its significant blue asymmetries for many quasars. 10\,438 of the 24\,357 quasars here are included in the sample used in \citet{Rankine2020}. The majority of the additional quasars possess spectrum S/N below the threshold of five imposed for the \citet{Rankine2020} sample. At such low S/N, redshift determinations using the reconstruction scheme can possess large errors. Systemic redshifts were therefore estimated using a set of 27 quasar templates spanning the full range of morphologies of the \CIII-emission complex shown in fig. A2 of \citet{Rankine2020}, which, again, deliberately exclude the \CIV-emission line. The analysis presented in the rest of this paper uses the quasars with S/N $\geq5$. The analysis was also repeated using the sample over the full spectrum S/N range and the results of the paper were unchanged.

Once systemic redshift estimates are available, the quasar spectra were reconstructed using the same ICA-scheme presented in section 4 of \citet{Rankine2020}. See Fig.~\ref{fig:spec} for two example spectra and their reconstructions. The spectrum-reconstructions essentially eliminate the uncertainties resulting from the presence of absorption features coincident with the \CIV-emission. The effective S/N of the reconstructions is also significantly improved relative to the observed spectra. \CIV- and \HeII- emission parameters and bolometric luminosities were calculated from the spectrum reconstructions. In particular, bolometric corrections $\text{BC}_{3000}=5.15$ and $\text{BC}_{1350}=3.81$ from \citet{Shen2011} were applied, as appropriate, to the rest-frame 3000\,\AA\ monochromatic luminosity or to the 1350\,\AA\ luminosity if 3000\,\AA\ is not available in the spectra.


The S/N $\geq5$ quasar sample contains 10\,547 quasars with redshifts $1.5 \leq z \leq 3.5$ from the SDSS DR14Q catalogue within the $\simeq$400\,deg$^2$ of LDR1. $\simeq$96 per cent of the quasars (10\,163) have reliable reconstructions and \CIV\ measurements based on the criteria detailed in \citet{Rankine2020}. \hl{1662 of the quasars have $\geq$5$\sigma$ LOFAR detections using the peak flux densities from the value-added catalogue of} \citet{Williams2019}. \hl{1528 of the 1662 detected quasars ($\simeq$92 per cent) possess reliable ICA-based reconstructions.} Our final quasar sample is formed by the 10\,163 quasars with S/N $\geq5$ and reliable reconstructions. A summary of the quasar sample is presented in Table~\ref{tab:samp}. 

BAL quasars make up only $\simeq$10 per cent of our sample and small-number statistics preclude an effective investigation of the radio and ultraviolet emission properties of the BAL quasars alone. A key result of the \citet{Rankine2020} paper was the high degree of similarity in the ultraviolet emission properties of the BAL and non-BAL quasar populations. The comparison of the two populations was possible because of the effectiveness of the spectrum reconstruction scheme described in \citet{Rankine2020}. We therefore use the spectrum reconstructions for the combined populations in our analysis, although none of the conclusions presented in the paper change if the analysis is confined to the non-BAL quasar population. 


\begin{table}
 \caption{S/N $\geq5$ quasar sample}
 \label{tab:samp}
 \begin{tabular}{lrl}
     \hline
     & Number & Fraction \\
     \hline
     All quasars & 10\,547 & 1\\
     LDR1 detected & 1662 &  0.16 \\
     LDR1 detected, reliable recons. & 1528 & 0.14 \\
     \hline
 \end{tabular}
\end{table}

\section{Results}
\label{sec:results}

\subsection[{Radio detected sources in C IV emission space}]{Radio detected sources in C\,{\sevensize IV} emission space} 
\label{sec:CIVres}

In this section, we present the distribution of LOFAR-(un)detected quasars in the plane of \CIV\,$\lambda$1549 equivalent width (EW) against blueshift; hereafter referred to as the \CIV\ emission space. The \CIV\ parameters are calculated in the same manner as in \citet{Rankine2020} and we refer the reader to their section 6.1 for a detailed description of the procedure. In summary, a non-parametric approach 
is taken whereby the EW of the line is calculated by integrating the continuum-subtracted spectrum reconstructions relative to the power-law continuum. The use of the reconstructions means that there is no requirement for fitting Gaussian profiles to the spectra. The blueshift of the line is calculated from the flux-weighted median wavelength of the \CIV\ emission relative to the systemic velocity of the quasar. The \CIV\ \hl{blueshift is often used to infer the strength of outflowing winds emanating from the accretion disc} \citep[e.g.][]{richards_unification_2011} \hl{with quasars with large (positive in this paper) blueshifts having strong winds.}

\begin{figure}
\includegraphics[width=\columnwidth]{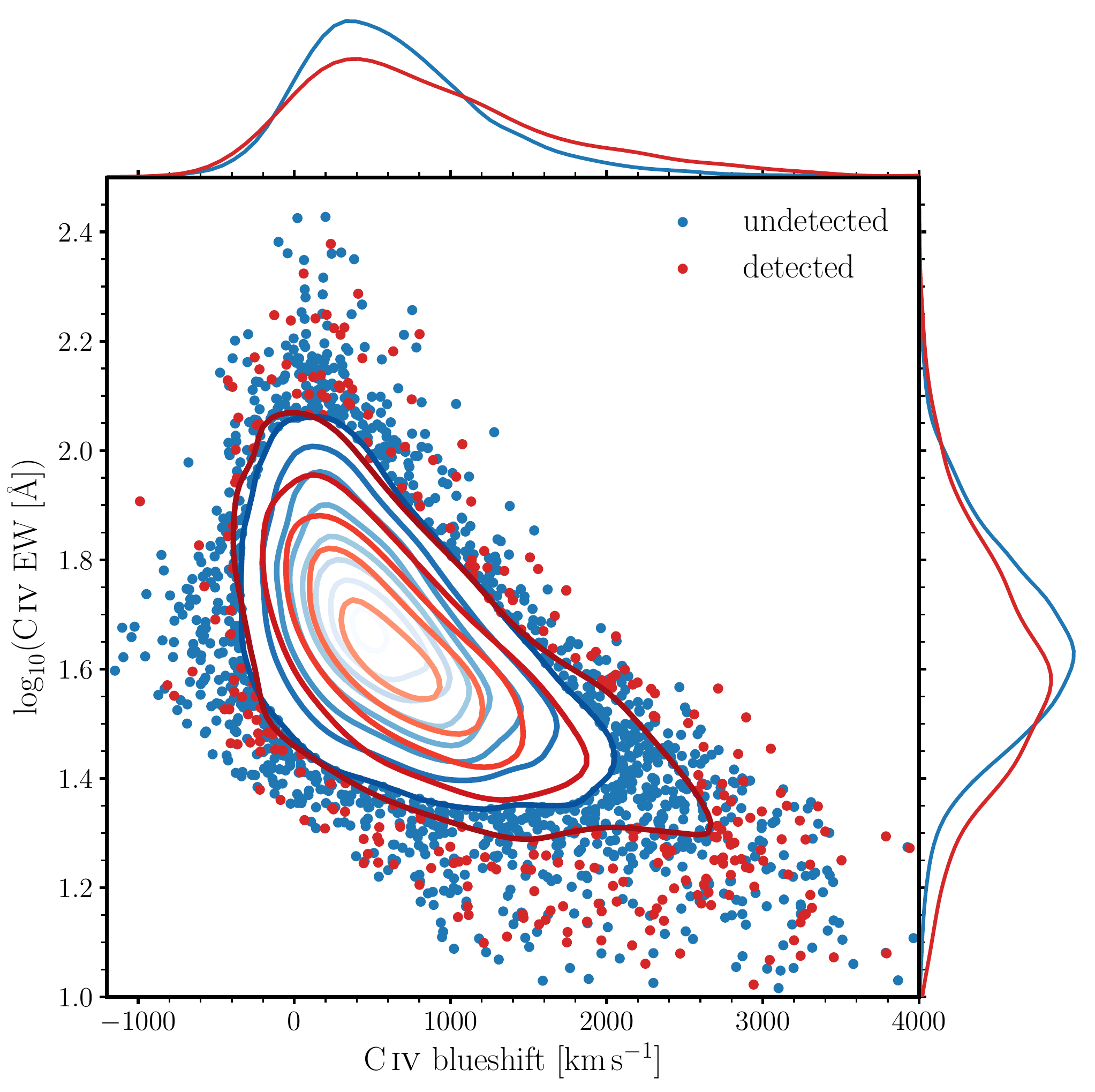}
\caption{Distribution of quasars detected (red dots/contours) and undetected (blue dots/contours) by LOFAR in \CIV\ emission space. \hl{The detected and undetected quasars populate the same region of {\ion{C}{iv}} space.}}
\label{fig:CIVdet}
\end{figure}

\begin{figure}
    \centering
    \includegraphics[width=\linewidth]{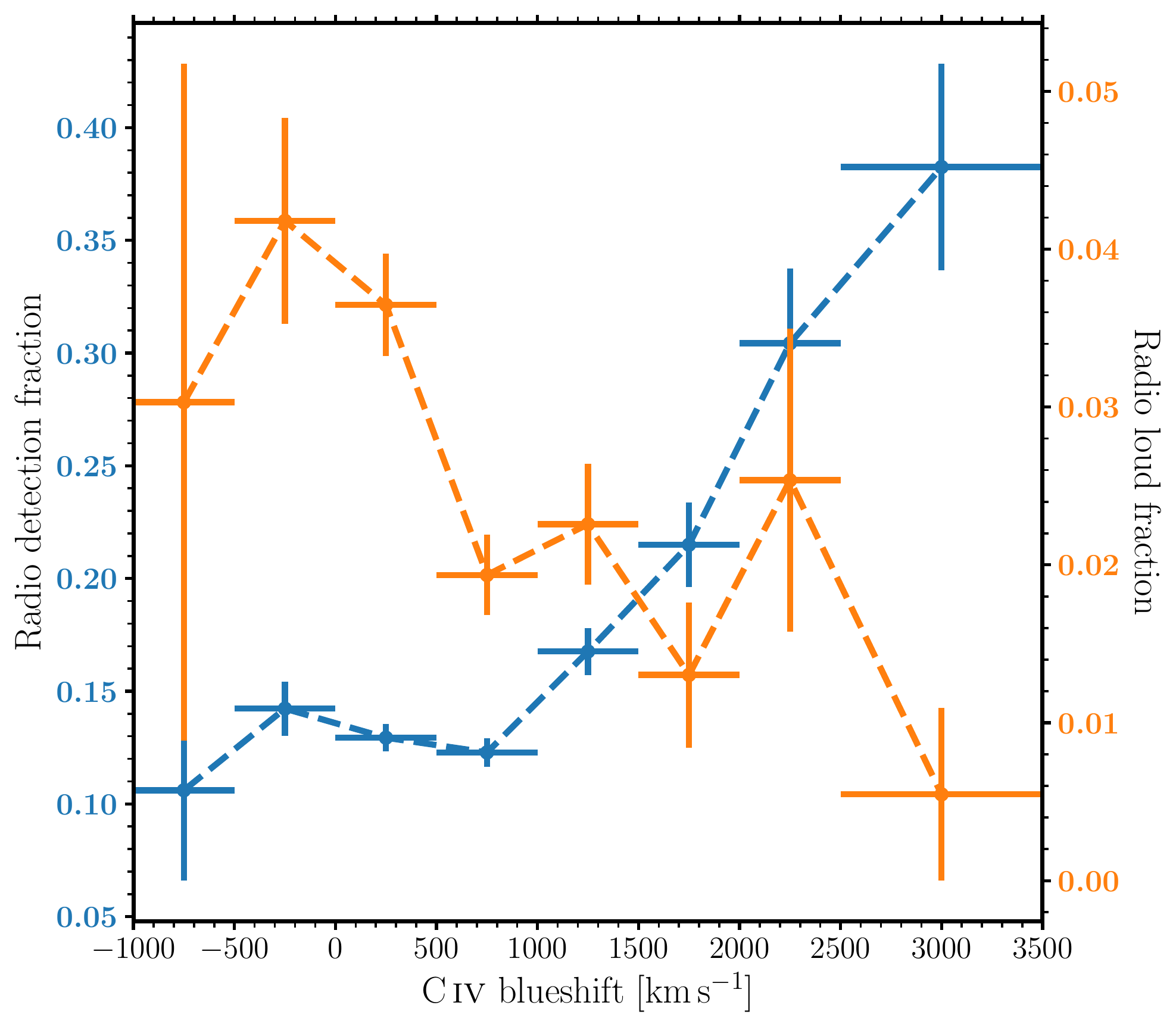}
    \caption{Radio-detection fraction (blue; left axis) and radio-loud fraction (orange; right axis) as a function of \CIV\ blueshift in bins of 500\kms, with the two highest blueshift bins combined. 1$\sigma$ uncertainties are calculated using Poisson errors. Note that the scales of the left and right axes are different. While the radio-detection fraction increases with blueshift, the radio-loud fraction decreases but with a more shallow gradient.}
    \label{fig:rdf}
\end{figure}

The distributions of LOFAR-detected and undetected quasars in \CIV\ emission space are shown in Fig.~\ref{fig:CIVdet}. The general distribution of all quasars is evident in this figure: quasars can either have strong \CIV\ emission (large EW) \textit{or} strong \CIV\ outflows (large blueshifts), but not both. However, quasars with weak emission and no outflow signature also exist. Note that the sharp diagonal line in Fig.~\ref{fig:CIVdet} is the result of excluding quasars in the sparsely populated low-blueshift, low-EW corner of \CIV\ emission space\footnote{Quasars in this area of \CIV\ emission space are typically FeLoBALs, have pathological spectra or have suboptimal reconstructions \citep[see section 4.3 of][for details]{Rankine2020}.}. Upon dividing the sample into quasars with and without LOFAR detections (see Section~\ref{sec:data}), \hl{it is evident that LOFAR-detected and undetected quasars can be found across the same region of {\ion{C}{iv}} emission space.} However, Fig.~\ref{fig:rdf} shows that the radio-detection fraction increases with blueshift from $\simeq$12 per cent at $\simeq0$\kms\ to $\simeq$40 per cent at $\simeq3000$\kms.

Is the correlation between radio-detection fraction and \CIV\ blueshift a selection effect? Our quasar sample is dominated by objects targeted in SDSS-III/BOSS and in SDSS-IV/eBOSS. SDSS-III/BOSS's target selection maximized the surface density of quasars at $z>2.15$ \citep{Ross2012} while the SDSS-IV/eBOSS target selection was designed to achieve a surface densities of $\sim$70\,deg$^{-2}$ quasars for $0.9<z<2.2$, $\sim$3--4\,deg$^{-2}$ for $z>2.1$ and included targets with FIRST detections \citep{Myers2015}. Our sample includes quasars at $1.5 \leq z \leq 3.5$ and there is a mild correlation between \CIV\ blueshift and (cosmological) redshift. However, we are cautious not to quantify any evolution of the detection fraction with redshift owing to potential selection effects resulting from the different target selections adopted by SDSS-III and -IV \hl{(but see} Appendix~\ref{app:redshift} \hl{where we show that the detection and radio-loud fractions as a function of {\ion{C}{iv}} blueshift do not change qualitatively with redshift).} We also test excluding the 14 quasars targeted purely for their FIRST detections as well as limiting the sample to only BOSS quasars targeted as part of the CORE sample and observe no qualitative changes in the detection or radio-loud fractions as a function of \CIV\ blueshift (see Appendix~\ref{app:selection}).

\begin{figure}
    \centering
    \includegraphics[width=\columnwidth]{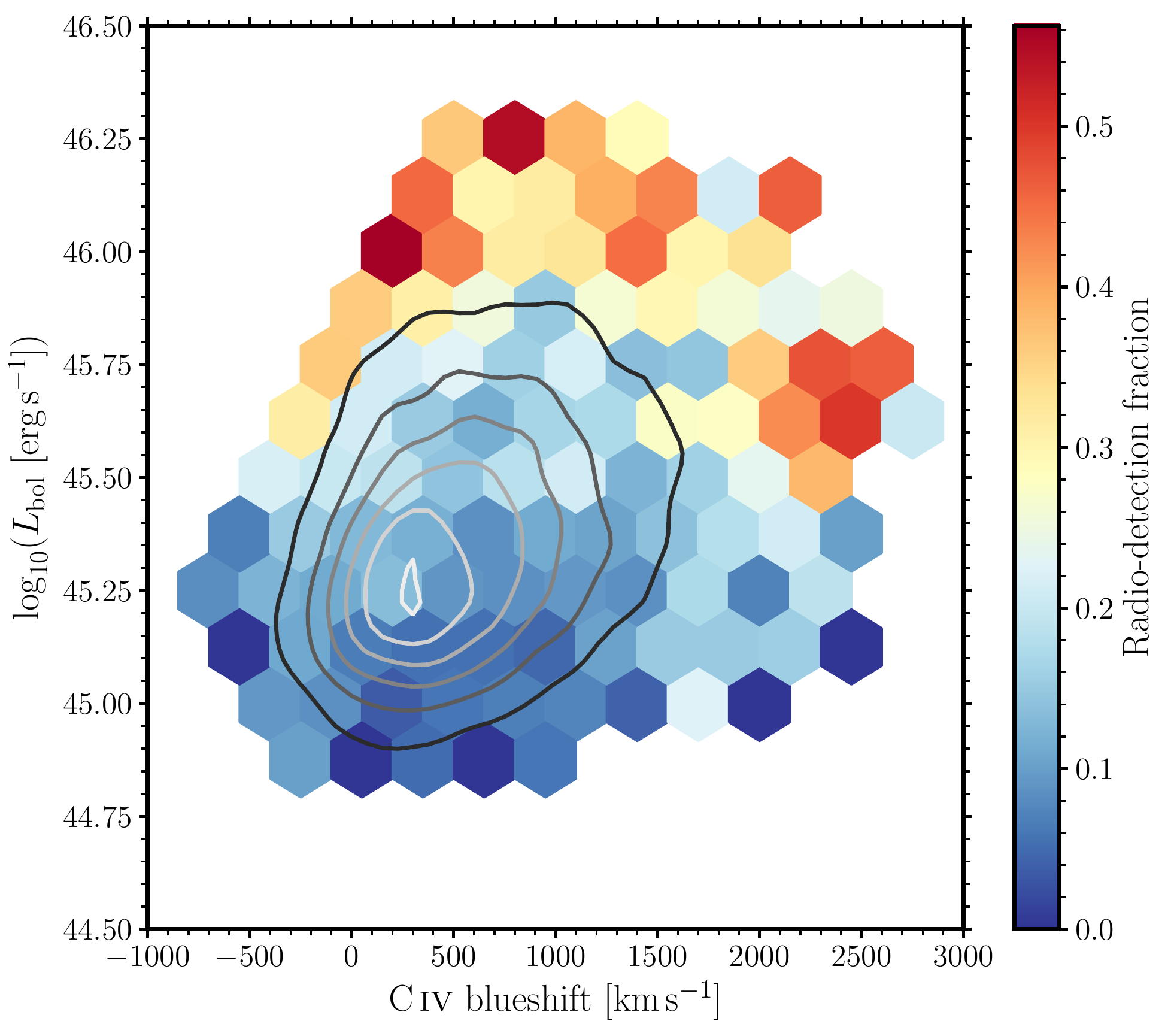}
    \caption{Radio-detection fraction as a function of \CIV\ blueshift and bolometric luminosity. Bins with 10 or more quasars are plotted and the contours illustrate the distribution of objects in \CIV\ blueshift-\Lbol\ space. \hl{The detection fraction increases with blueshift and} \Lbol. \hl{While this trend is stronger with the latter, there is still a trend of increasing detection fraction with blueshift at constant} \Lbol.}
    \label{fig:RDFhex}
\end{figure}

Conversely, \CIV\ blueshift is known to increase with bolometric luminosity \citep[see contours in Fig.~\ref{fig:RDFhex} and][]{Rankine2020} and the radio-detection fraction also increases with \Lbol. However, Fig.~\ref{fig:RDFhex} illustrates that the increase of radio-detection fraction is empirically found with both \Lbol\ and blueshift. The trend of increasing radio-detection fraction with \Lbol\ is stronger than with \CIV\ blueshift but at fixed luminosity there is still a tendency for the radio-detection fraction to increase with blueshift. We have checked that the dependence of \CIV\ blueshift on radio-detection fraction is still apparent when restricting the sample to quasars with $\log_{10} L_{\text{bol}} < 45.5$. However, at luminosities $\log_{10} L_{\text{bol}} > 45.75$ the detection fraction first decreases and then increases with increasing blueshift. This change in the behaviour of the detection fraction at low blueshifts is a result of removing quasars with large \CIV\ EWs from the sample when excluding quasars with low \Lbol\ \citep[see fig. 13 of][]{Rankine2020} thus preferentially removing undetected quasars at low blueshifts (see Fig.~\ref{fig:CIVdet}). Hinted at in Figs.~\ref{fig:CIVdet} and \ref{fig:RDFhex} is that the radio detection trend with blueshift at fixed EW is likely more complex than the simple monotonic increase observed in the whole sample. It should be highlighted that in this paper we are investigating a few key variables, namely \CIV\ blueshift, \HeII\ EW, and bolometric luminosity, that are providing particular projections in what is likely a multidimensional parameter space.

\subsection{The radio-loud fraction}
Radio loudness, defined as 
\begin{equation}
    R = \frac{L_{\text{rad}}}{L_{\text{opt}}},
\end{equation}
with radio luminosity $L_{\text{rad}}$ and optical luminosity $L_{\text{opt}}$, has often been used as a proxy for dividing radio sources into two populations: radio-loud sources have powerful, large-scale Fanaroff-Riley type jets \citep{Fanaroff1974}, and radio-quiet sources have other radio emission mechanisms (which can include star formation). 

A threshold that minimizes the overlap in each sample can be used as a crude cut to separate radio-quiet from radio-loud sources, although with contamination in both samples. An appropriate value is often identified from a dip in the overall distribution of radio loudness, and the classical radio-loud threshold is $\log_{10}R = 1$ at 5\,GHz \citep{kellermann_vla_1989}. \hl{We have extrapolated this threshold to 144\,MHz using the typical synchrotron spectral index of $-$0.7 to define a radio-loud threshold of $\log_{10} R = 2$ for our sample.}

\begin{figure}
\includegraphics[width=\columnwidth]{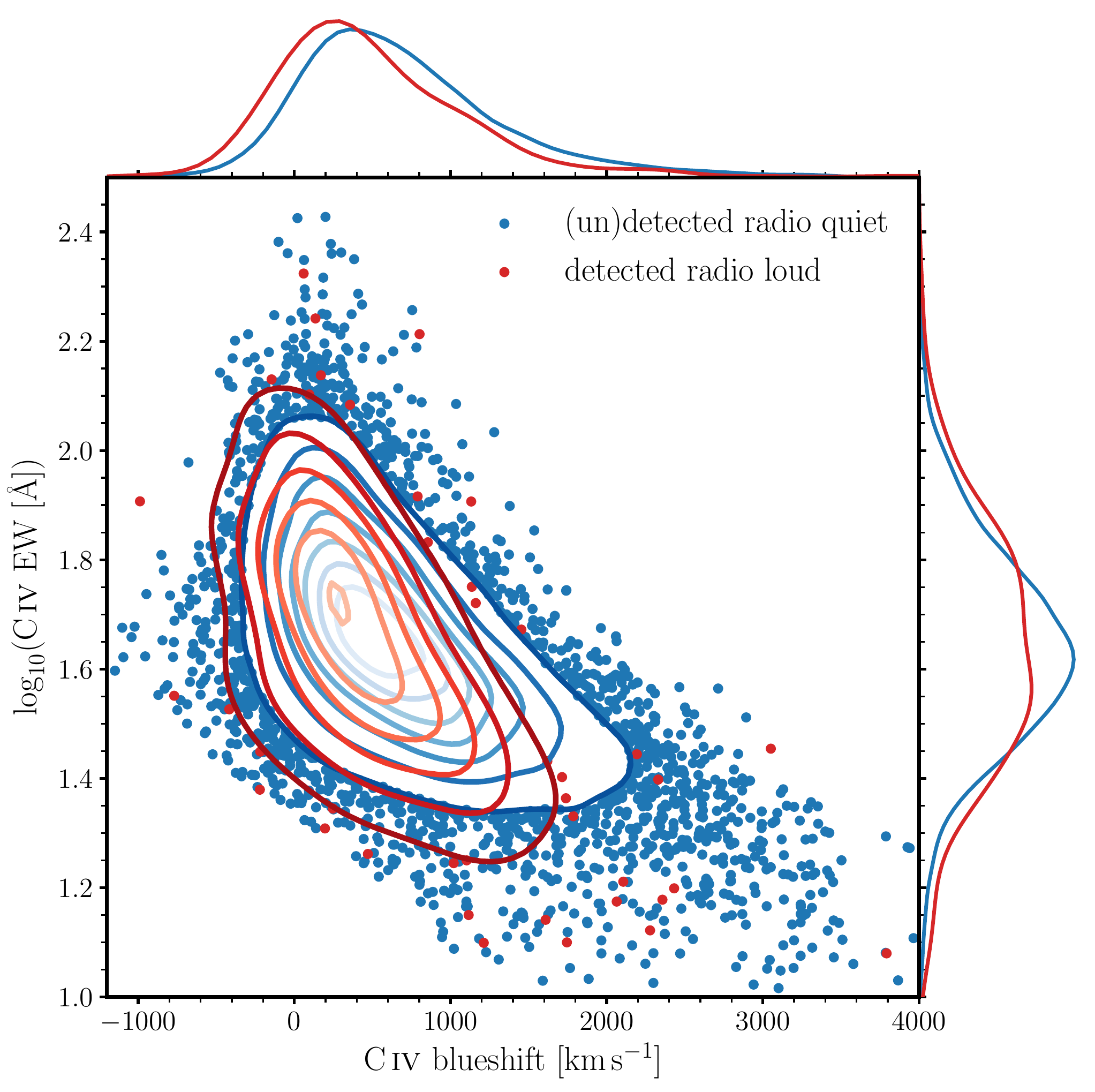}
\caption{Radio-loud (red dots/contours) and radio-quiet (blue dots/contours) quasars in \CIV\ emission space. \hl{Radio-loud and radio-quiet quasars can be found in the same region of {\ion{C}{iv}} space that the undetected quasars populate but the radio-loud quasars are skewed towards lower {\ion{C}{iv}} blueshifts than the radio-quiet sources.}}
\label{fig:CIVRloud}
\end{figure}

In Fig.~\ref{fig:CIVRloud}, we again show the \CIV\ emission space, now dividing the quasar sample into radio-loud and radio-quiet sources. Here we use the SDSS $i$-band magnitude, corresponding to rest-frame 2500\,\AA\ for the median redshift of the sample, to calculate the optical luminosity. Radio luminosities are calculated from the integrated flux densities in LDR1 and utilise the ICA- and template-derived spectroscopic redshifts described in Section~\ref{sec:data}. Both radio and optical luminosities are k-corrected to $z=0$ using spectral indices $-$0.7 and $-$0.5, respectively. We experimented with different radio-loud thresholds and although the results change quantitatively, the conclusions of the paper are unchanged. In addition to the detected radio-quiet quasars, we also included the undetected radio-quiet sources by estimating the upper limit on \logR\ by calculating the radio luminosity at the flux limit of 0.35\,mJy for the undetected objects. 8631 undetected objects with an upper limit of $\log_{10} R < 2$ are included in the radio-quiet population for the remainder of this subsection. \hl{Four undetected quasars have upper limits of $\log_{10} R > 2$ with the potential to be radio-loud thus are not included in either sample.} While radio-loud quasars are present at all locations in \CIV\ emission space, they are concentrated at low \CIV\ blueshifts ($<500$\kms). The radio-loud fraction (RLF) is calculated from 
\begin{equation}
    \text{RLF} = \frac{N_{\text{RL,det}}}
        {N_{\text{RL,det}} + N_{\text{RQ,det}} + N_{\text{RQ,undet}}},
\end{equation}
where the denominator includes the number of undetected radio-quiet quasars ($N_{\text{RQ,undet}}$) as well as the number of detected radio-quiet ($N_{\text{RQ,det}}$) and radio-loud ($N_{\text{RL,det}}$) quasars. The radio-loud fraction as a function of \CIV\ blueshift is presented in Fig.~\ref{fig:rdf} and is shown to decrease from around 4--5 per cent to $<1$ per cent as blueshift increases from $\simeq0$ to $\simeq3000$\kms, in qualitative agreement with \citet{richards_unification_2011} and \citet{kratzer_mean_2015} using FIRST.

\begin{figure}
    \centering
    \includegraphics[width=\columnwidth]{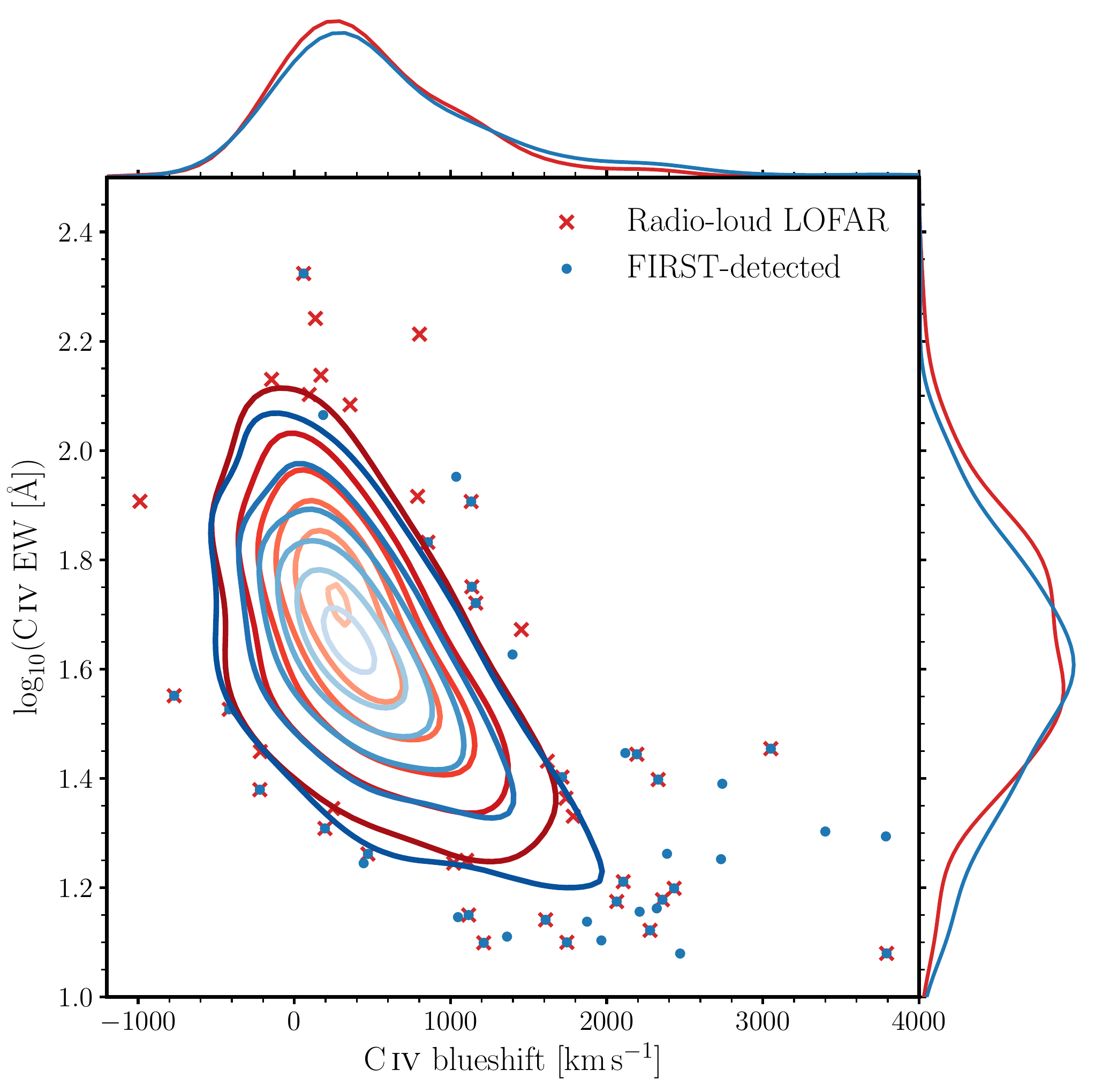}
    \caption{The distribution of FIRST-detected sources (blue dots/contours) is identical to the distribution of radio-loud sources in LOFAR (red crosses/contours; same as in Fig~\ref{fig:CIVRloud}) in \CIV\ emission space. All FIRST-detected sources are also detected in LOFAR unless they have inverted spectra with $\alpha \gtrsim 0.46$.
    }
    \label{fig:FIRST}
\end{figure}

The design goal of FIRST was to achieve sufficient sensitivity to detect the star-forming galaxy population below the break in the radio logN-logS curve \citep[see][section 3.2]{Becker1995}. For 5$\sigma$ detections, FIRST has a flux limit of 1\,mJy at 1.4\,GHz, as compared to LOFAR’s 0.35\,mJy at 144\,MHz \citep{Rosario2020}. As a result, FIRST is mostly sensitive to the radio-loud population, and all FIRST-detected sources are also detected in LOFAR unless they have inverted spectra with $\alpha \gtrsim 0.46$. Not all LOFAR-detected, radio-loud sources are detected in FIRST, due to the combination of sensitivity and observing frequency — such sources can either be fainter overall, or sufficiently steep spectrum such that they could not be detected by FIRST. We have matched the SDSS quasars that are in the LOFAR footprint to FIRST using the nearest-neighbour LOFAR--FIRST matching procedure described in section 2.4 of \citet{Morabito2019}. Figure~\ref{fig:FIRST} demonstrates that the radio-loud sources that are detected in LOFAR have an extremely similar distribution in \CIV\ emission line space to the FIRST-detected sources. This similarity suggests that the LOFAR-detected radio-loud sources are an extension of the FIRST-detected population studied by \citet{richards_unification_2011}, even though these sources have fainter radio emission at 1.4\,GHz compared to FIRST-detected objects. Equivalently, it implies that, when a reasonable radio-loudness cut is applied, the two observing frequencies are tracing similar AGN phenomena since the location in \CIV\ emission space is predicated upon the properties of the quasar which could be related to the origin of the radio emission.

\subsection[{Radio properties in C IV emission space}]{Radio properties in C\,{\sevensize IV} emission space}
\label{sec:properties}

\begin{figure*}
    \centering
    \includegraphics[width=\linewidth]{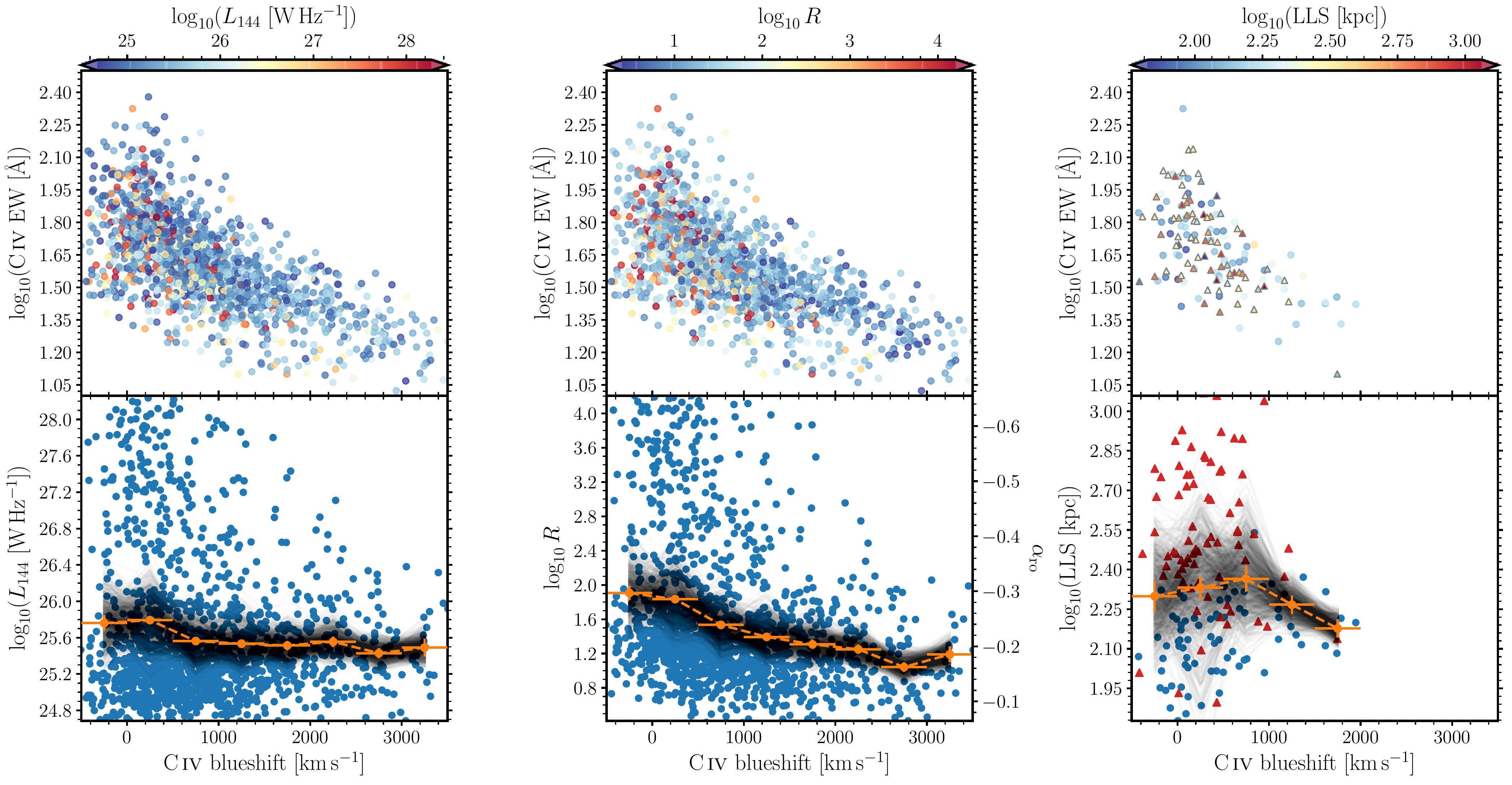}
    \caption{Top: \CIV\ emission space for detected sources with points coloured, from left to right, by $\log_{10}L_{144}$, \logR\ and (resolved) largest linear size. Bottom: The respective radio property of the detected sources in the top panels against \CIV\ blueshift (blue points). Note that the middle panel with \logR\ on the left $y$-axis has the corresponding $\alpha_{\text{ro}}$ values on the right $y$-axis. Quasars with LOFAR Galaxy Zoo sizes (see text for details) are plotted as triangles in the right-hand panels (and red in the bottom panel). The mean log radio property in bins of width 500\kms\ are plotted in orange with the standard error on the mean. The grey lines are 1000 `bootstrapped curves': the number of objects in each bin is controlled by randomly choosing $N$ objects per bin, where $N$ is the number of objects in the highest \CIV\ blueshift bin, and plot the mean.}
    \label{fig:multiple}
\end{figure*}

Figure~\ref{fig:multiple} contains the \CIV\ emission space populated by the LOFAR-detected quasars, revealing trends with blueshift for various radio properties. The radio luminosity (left-hand panels) reveals that low-luminosity radio emission can be found in quasars across the \CIV\ emission space but quasars with high-luminosity radio emission typically exist at the low-blueshift end of observed \CIV\ emission profiles. We ascertain that the lack of radio-bright quasars with high blueshifts is not an artefact of fewer quasars overall in this region of \CIV\ emission space by performing bootstrap sampling in bins of width 500\kms, where the number of samples in each bin is controlled by the number of objects in the highest \CIV\ blueshift bin. The mean log radio luminosity of the samples in each bin is calculated. The bootstrapping is repeated 1000 times, producing the 1000 grey curves in the lower left-hand panel of Fig.~\ref{fig:multiple}. The curves are consistent with no maximally-luminous sources at high blueshift and in qualitative agreement with the trend in the mean radio luminosity of decreasing $L_{144}$ with increasing blueshift in the whole population. Note that the calculation of the mean and bootstrapping is performed using the log of the radio parameters, but conclusions based on the panels in Fig.~\ref{fig:multiple} are unchanged if the analysis is performed on the linear radio parameters.

The bolometric luminosity is known to increase with blueshift \citep[see Fig.~\ref{fig:RDFhex} and][]{Rankine2020}. However, the decrease in the mean of \logR\ with blueshift (middle panels of Fig.~\ref{fig:multiple}) demonstrates that quasars with little to no outflowing component of \CIV\ are, on average, more radio-luminous relative to their optical luminosity than the quasars with a significant outflowing component. 
\hl{We have verified that this correlation with blueshift holds even after taking the increasing average optical luminosity into account, as would be expected from the fact that the mean radio luminosity also decreases with blueshift (left-hand panel of} Fig~\ref{fig:multiple}).
Again, the bootstrap sampling of \logR\ illustrates the decreasing average radio-loudness with increasing blueshift \citep[consistent with][]{kratzer_mean_2015} is not caused by fewer quasars at high blueshifts. Alongside \logR, we calculate \aro, the radio-to-optical spectral index, defined as
\begin{equation}
    \alpha_{\text{ro}} = \frac{\log_{10}(L_{\text{rad}} / L_{\text{opt}})}{\log_{10}(\lambda_{\text{opt}} 
    / \lambda_{\text{rad}})} 
    = \frac{\log_{10}R}{\log_{10}(\lambda_{\text{opt}} 
    / \lambda_{\text{rad}})}, 
\end{equation}
with radio luminosity, $L_{\text{rad}}$, at $\lambda_{\text{rad}}=2$\,m, and optical luminosities derived from SDSS $i$-band photometry ($\lambda_{\text{opt}}=7625$\,\AA). The radio-loud threshold of $\log_{10} R = 2$ is equivalent to $\alpha_{\text{ro}} \simeq-0.31$, with more radio-loud objects having more negative \aro\ values. We label the right-hand axis of the lower panel of the middle column of Fig.~\ref{fig:multiple} with the equivalent \aro\ values for \logR.

The existence of at least two populations is clear from the bottom left and middle panels of Fig.~\ref{fig:multiple}. There is a clear division at $L_{144} = 10^{26}$\,W\,Hz$^{-1}$ across all \CIV\ blueshifts. On the contrary, a division in radio-loudness appears to be dependent on \CIV\ blueshift suggesting that the radio-loud threshold should decrease with blueshift rather than being a constant value. We choose not to adopt a blueshift-dependent radio-loud threshold in order to be consistent with previous papers.

In the right panels of Fig.~\ref{fig:multiple}, we present the projected largest linear size (LLS) of the resolved radio emission. Where available, the LOFAR Galaxy Zoo sizes \citep{Williams2019} are used and are presented as triangles (and red in the lower right panel). By design, the quasars selected for analysis with Galaxy Zoo typically have large radio sizes. All other radio sizes are calculated from twice the full width at half-maximum of the deconvolved major axis [see section 2.1 of \citet{Hardcastle2019} for why the doubling is required] presented as (blue in lower right panel) circles. There is a trend of decreasing LLS with increasing blueshift, and, at high blueshifts, there are no sources large enough to be resolved by LOFAR.

\begin{figure}
    \centering
    \includegraphics[width=\linewidth]{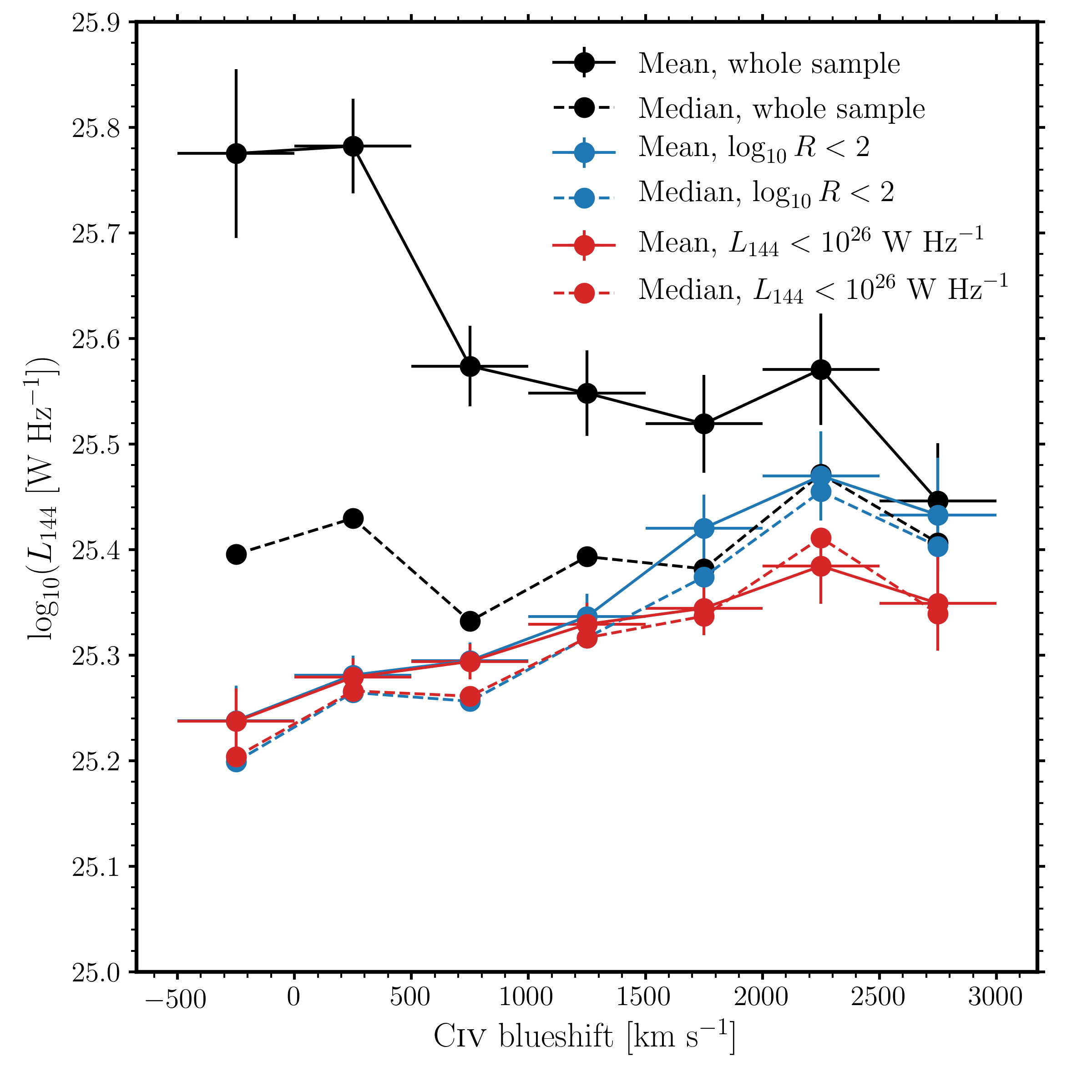}
    \caption{The mean and median values of $\log_{10} (L_{144})$ as a function of blueshift, using the same 500\kms\ bins as in Fig.~\ref{fig:multiple}, with various cuts applied to the data. Error bars show the bin size and the standard error on the mean. The mean and median from the whole sample is shown in black; the mean black curve is identical to that in the bottom left panel of Fig.~\ref{fig:multiple}. For the whole sample, the mean curve decreases with \CIV\ blueshift as in Fig.~\ref{fig:multiple}, whereas the median marginally increases. The red and blue curves show the mean and median $\log_{10} (L_{144})$ for sources with $\log_{10} R < 2$ (blue) and $L_{144} < 10^{26}\,{\rm W\,Hz}^{-1}$ (red). The radio-quiet and low-luminosity sources evolve differently across \CIV\ emission line space to the luminous, radio-loud population.
    }
    \label{fig:l144_cuts}
\end{figure}

In Fig.~\ref{fig:multiple}, we plotted mean radio luminosity in bins of \CIV\ blueshift, but the mean can be misleading since there are multiple radio populations contributing. The black points in Fig.~\ref{fig:l144_cuts} show the mean and median values of $\log_{10} (L_{144})$ in bins of blueshift for the whole sample, showing a decreasing mean $\log_{10} (L_{144})$ and a flat or perhaps marginally increasing median $\log_{10} (L_{144})$ as \CIV\ blueshift increases. Also in Fig.~\ref{fig:l144_cuts}, we plot mean and median $\log_{10} (L_{144})$ against \CIV\ blueshift after applying cuts to the sample of $\log_{10} R < 2$ and, separately, $L_{144} < 10^{26}$\,W\,Hz$^{-1}$ (red and blue points). The cut in \logR\ removes the radio-loud population shown in the middle panels of Fig.~\ref{fig:multiple}. The luminosity cut removes mostly radio galaxies with powerful jets; however, less powerful jets do exist down to $L_{144} \simeq 10^{22}$\,W\,Hz$^{-1}$ \citep[see fig. 5 of][]{Mingo2019}. Both cuts produce increasing mean and median $L_{144}$ with blueshift illustrating that the low-luminosity and radio-quiet quasars correlate with blueshift in a different manner from the luminous and radio-loud population. Different correlations for different populations is consistent with the differing trends in radio-detection and radio-loud fractions observed in Fig.~\ref{fig:rdf}.

\subsection[{He II properties and size-luminosity diagrams}]{He\,{\sevensize II} properties and size-luminosity diagrams}

The strength of the \HeII\,$\lambda1640$ recombination line is indicative of the strength of the soft X-ray SED \citep{Leighly2004}. The strong anticorrelation between \CIV\ blueshift and \HeII\ EW in \citet{Rankine2020} \citep[see also][]{Baskin2013, Baskin2015} suggests that winds can only be launched in quasars which have weak soft ionizing SEDs such that material can be accelerated by line driving owing to electrons remaining bound to the nuclei. In this section, we investigate the relationship between the strength of the \HeII\ line and the radio emission.

Fig.~\figheiilum\ (left-hand panel) illustrates the relationship between the \HeII\ EW and the radio and bolometric luminosities. There are two populations in \Lbol-$L_{144}$ space: the radio-loud (triangles) and radio-quiet (circles) quasars. The radio-quiet quasars follow the trend of increasing radio luminosity with increasing bolometric luminosity, which is by definition for the upper envelope of the radio-quiet quasars since \logR\ sets the dividing line between loud and quiet sources. However, the lower envelope of the radio luminosity is also increasing with \Lbol, which, if not a selection effect, suggests that \logR\ should be favoured over radio luminosity for defining the radio-loud population \citep[see][]{Balokovic2012}. There is also evidence for decreasing SED hardness via the decrease in \HeII\ EW as \Lbol\ increases. The anticorrelation between \Lbol\ and \HeII\ EW is not unlike the Baldwin effect \citep{Baldwin1977}. As expected from the radio-loud definition, the radio-loud quasars are scattered to higher radio luminosities and show no evidence for a trend with \Lbol. Note there is not a sharp dividing line between the radio-loud and radio-quiet populations. \logR\ is calculated using $L_{2500}$, which is determined from the SDSS photometry, while \Lbol\ is calculated from the spectra. The radio loud quasars typically have strong \HeII\ emission, indicating a hard ionizing continuum, which could suggest that jets and SED shape are correlated with the same underlying physics (see Section~\ref{sec:rlf_trend}).

\begin{figure*}
    \centering
    \includegraphics[width=\linewidth]{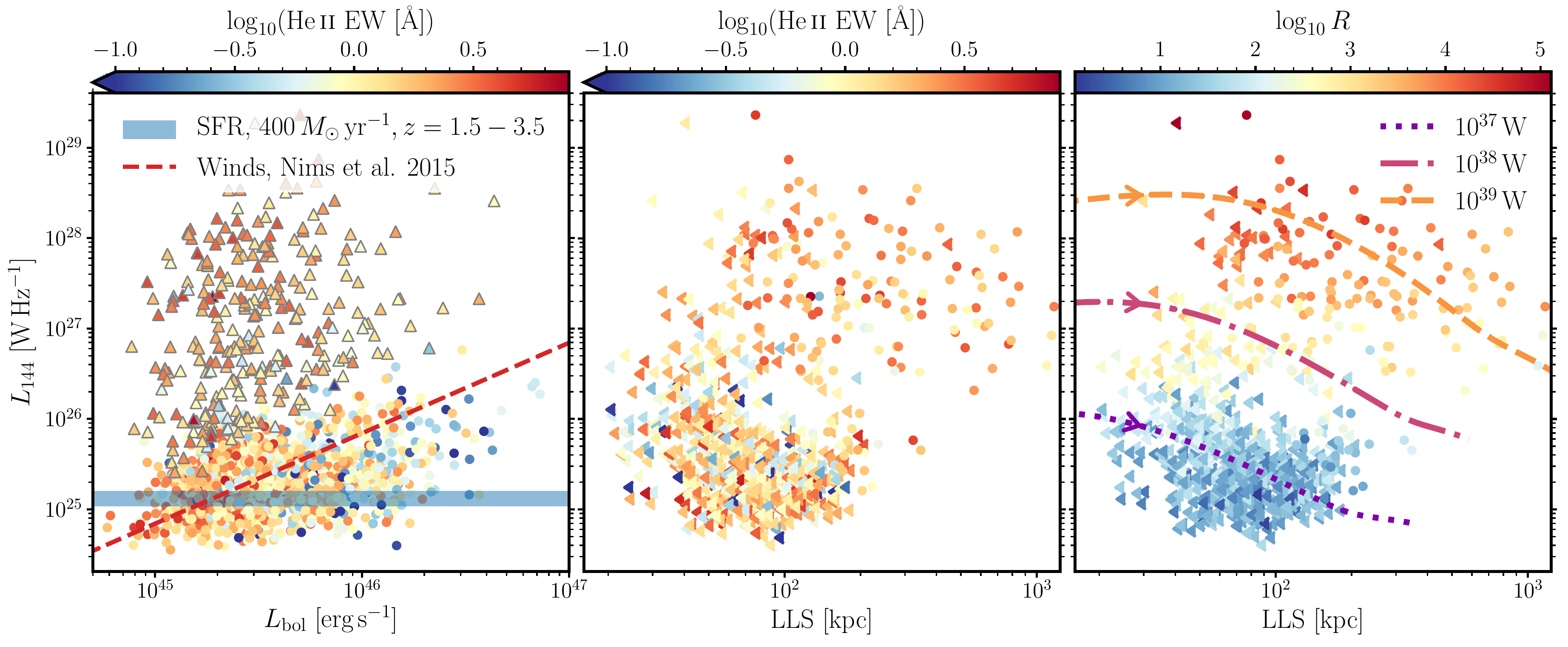}
    \caption{Left: Radio luminosity against bolometric luminosity. Circles (radio-quiet) and triangles (radio-loud), all coloured by $\log_{10}$(\HeII\ EW). In radio-quiet quasars, radio luminosity is correlated with bolometric luminosity, both of which are anticorrelated with \HeII\ EW, covering the full range in \HeII\ strength. The radio-loud quasars host more stochastic radio emission and favour strong \HeII\ emission. The red dashed line and blue band illustrate the expected radio emission from winds and star formation (see Sections~\ref{sec:SF} and \ref{sec:winds}). Middle: Radio luminosity against LLS. Circles are resolved sources and leftward arrows are unresolved. All unresolved LLS values are upper limits. Quasars with big/extended and bright radio sources typically have strong \HeII\ emission while small/compact and faint radio emission is hosted by quasars with a range of \HeII\ line strengths. Right: Same as middle but with circles (resolved) and leftward arrows (unresolved) coloured by \logR. Radio galaxy evolutionary tracks calculated from the models of \citet{Hardcastle2018} at $z=2$ are plotted.
    }
    \label{fig:heii_3panel}
\end{figure*}

In the middle panel of Fig.~\figheiipd, we plot the radio luminosity against LLS with points coloured by \HeII\ EW. This size--luminosity plot is often referred to as a P-D diagram and is useful for studying the evolution of sources as well as the origin of the radio emission \citep[e.g.][]{baldwin_evolutionary_1982,blundell_nature_1999,an_dynamic_2012,Hardcastle2019,hardcastle_radio_2020}. Again we see two clouds of points: quasars with relatively small ($\lesssim100$\,kpc) and faint radio emission ($\lesssim10^{26}$\,W\,Hz$^{-1}$), the majority of which are unresolved such that the sizes are upper-limits; and a second cloud of large and bright radio sources ($\gtrsim100$\,kpc and $\gtrsim10^{27}$\,W\,Hz$^{-1}$). The small-size, low-luminosity cloud covers the full range of \HeII\ EWs, while the large and bright cloud in the upper right is populated, almost exclusively, with quasars that have strong \HeII\ emission. The distribution of \HeII\ EW across the radio-loud and -quiet populations could be a sign that jets are semi-stochastic: they can form in high \HeII\ EW sources but are not required to (see Section~\ref{sec:stochastic}). The strong correlation between \CIV\ blueshift and \HeII\ EW identified in \citet{Rankine2020} means that Fig.~\figheiipd\ is qualitatively unchanged if we replace \HeII\ EW with \CIV\ blueshift, allowing it to be inferred that quasars with strong winds do not host bright, extended radio emission. \citet{stocke_radio_1992} arrived at this same conclusion using BALs as their evidence for winds. See also \citet{Mehdipour2019} who report an anticorrelation between the column density of ionized X-ray winds and the radio-loudness parameter.

We also show a size-luminosity diagram, with points coloured by $\log_{10} R$, in the right-hand panel of Fig.~\figtracks, together with radio galaxy evolutionary tracks. The tracks are calculated from the models of \citet{Hardcastle2018} for three different jet powers ($10^{37}$,$10^{38}$,$10^{39}$\,W) and evolved for $500$\,Myr in an environment defined using the universal pressure profile of \cite{arnaud_universal_2010}. We conduct the calculations at $z=2$ and adopt a cluster mass of $M_{500}=2.5\times10^{13}\,M_\odot$ with a temperature of $k_B T= 1$\,keV. These parameters are chosen so the tracks act as $z=2$ analogues to the $z=0$ tracks in figure 8 of \cite{Hardcastle2019}, and the aim is to give a feel for the size-luminosity evolution of jets with different powers in a group environment. The evolutionary tracks assume a jet-origin for the radio emission; however, we do not argue that this is the case. The right-hand panel of Fig.~\figtracks\ shows that radio-quiet quasars are not simply younger, compact versions of the jetted radio-loud quasars -- if the jet power is roughly constant. As sources get larger, their radio luminosity instead decreases for a given jet power, suggesting that any evolutionary link between the two populations would require a substantial change in jet power output. The tracks also show the typical jet powers required to produce the observed $L_{144}$ in radio-quiet and radio-loud sources (although caution should be employed when estimating jet powers from radio luminosity, see e.g. \citealt{hardcastle_radio_2020}). Taking the modelled jet powers at face value implies that radio-loud quasars require kinetic powers comparable to their total radiative outputs, whereas the radio-quiet sources can, unsurprisingly, be powered by much weaker jets.

\section{Discussion}
\label{sec:discussion}
Our investigation into the low-frequency radio and ultraviolet emission line properties of a sample of SDSS quasars allows for a discussion about the origin of the radio emission, the relative importance of jets, winds, and star formation and the physical drivers of the observed trends with \CIV\ and \HeII\ properties.

\subsection{Origin of Radio emission} 
Radio emission can come from multiple different sources. As discussed in the previous section, the radio-loud sources are typically large and luminous, whereas the radio-quiet quasars are often unresolved, suggesting fairly compact sources with low radio luminosities. It is well known that the radio emission in radio-loud quasars is dominated by jets; however, the origin of the radio in radio-quiet quasars is unknown and could be a combination of star formation, jets, and winds \citep[see ][for a review]{panessa_origin_2019}. In this section, we discuss the plausibility of these mechanisms as possible sources of the radio luminosities observed in our quasar sample, focusing specifically on the radio-quiet population.

\begin{figure}
    \centering
    \includegraphics[width=\columnwidth]{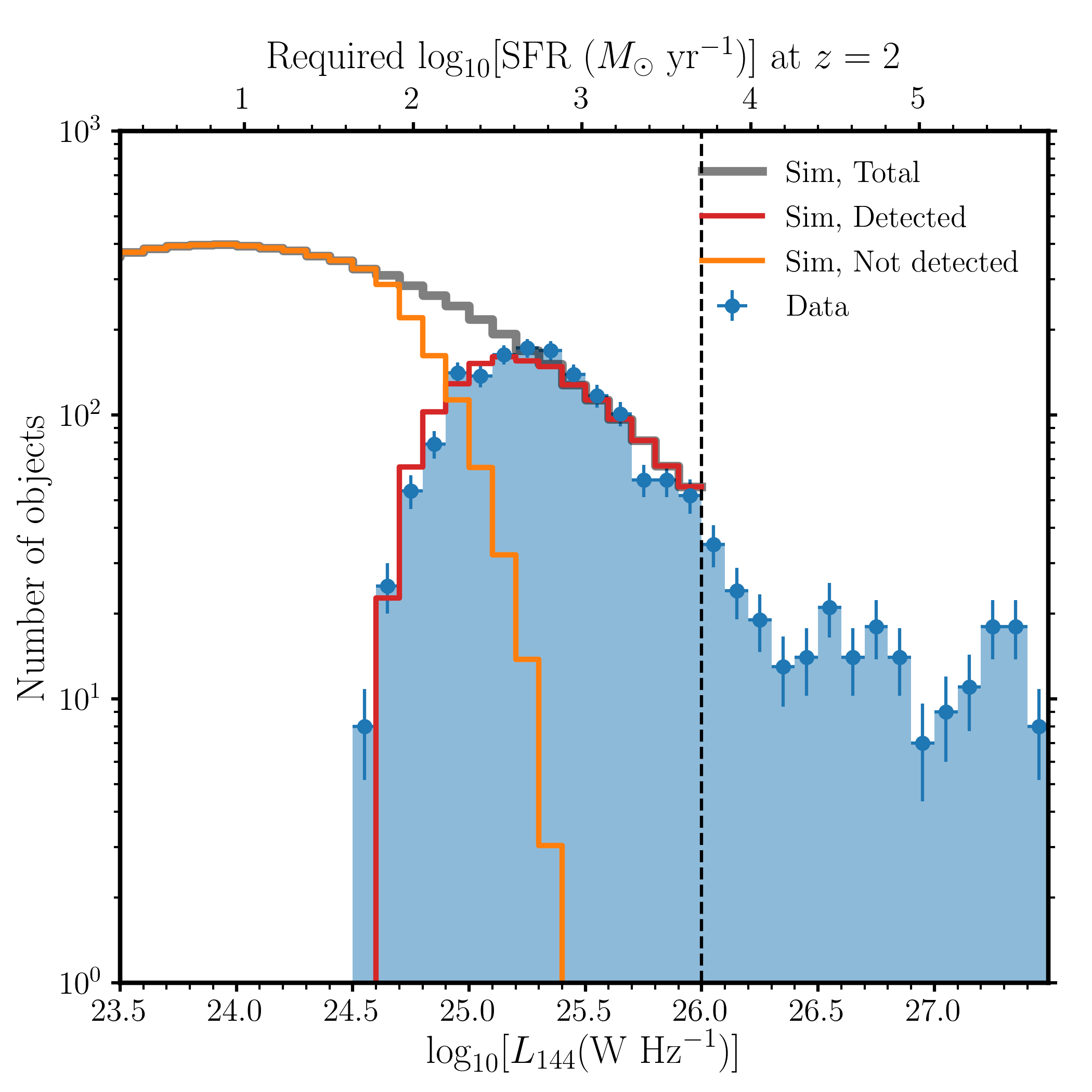}
    \caption{A histogram in radio luminosity, showing the sources from our sample compared to the simulated population from the Monte Carlo simulations described in Section~\ref{sec:SF}. The dotted line marks $L_{144}=10^{26}$\,W\,Hz$^{-1}$, above which sources were not included in the simulation (although the data are still plotted). The figure shows that star formation rates of $\sim300$ up to $1000$s of $M_\odot\,{\rm yr}^{-1}$ can plausibly produce the radio emission up to $L_{144} \sim 10^{26}$\,W\,Hz$^{-1}$. The limitations of the method, which is intended to be illustrative, are discussed briefly in the text. The relation to the radio luminosity function is discussed in Section~\ref{sec:contributions}.
    }
    \label{fig:SFR}
\end{figure}

\subsubsection{Star Formation}
\label{sec:SF}
Radio emission from star formation is produced by synchrotron radiation from non-thermal electrons accelerated in supernova remnant shocks, as well as free-free emission from \ion{H}{ii} regions ionized by massive stars \citep[see][for a review]{condon_radio_1992}. At the frequencies considered here, the non-thermal synchrotron component dominates. One way of calibrating the amount of radio emission produced by a given star formation rate (SFR) is from the far-infrared radio correlation \citep[FIRC;][]{helou_thermal_1985,yun_radio_2001}. Using spectral modelling combined with LOFAR data, \cite{calistro_rivera_lofar_2017} find the relationship 
\begin{equation}
    \frac{\mathrm{SFR}_{144}}{M_\odot\,{\rm yr}^{-1}} = 1.455 \times 10^{-24}\,10^{q(z)}~\frac{L_{144}}{\mathrm{W\,Hz}^{-1}}, 
    \label{eq:sfr}
\end{equation}
where $q(z)=1.72(1+z)^{-0.22}$ is a factor accounting for the redshift evolution of the FIRC. Using this relation, we can estimate the SFR needed to produce the radio emission in the cloud of radio-detected, radio-quiet quasars. These quasars have typical radio luminosities of $L_{144}\sim10^{25}$\,W\,Hz$^{-1}$, which from equation~\ref{eq:sfr} requires SFR$\sim 300\,M_\odot\,{\rm yr}^{-1}$ at $z=2$. This SFR is quite high, but not unreasonable for a quasar in our redshift range; for example, \cite{harris_star_2016} infer $\sim 300\,M_\odot\,{\rm yr}^{-1}$ at $z=2-3$ and \cite{stanley_mean_2017} find $\sim 50-250\,M_\odot\,{\rm yr}^{-1}$ at $z=1.5-2.5$. \cite{harris_star_2016} find their results are consistent with the main sequence of star-forming galaxies. Higher SFRs ($\gtrsim1000\,M_\odot\,{\rm yr}^{-1}$) have also been measured in both quasar hosts \citep{pitchford_extreme_2016} and radio galaxies \citep{drouart_rapidly_2014}. Inevitably, it is the most star-forming radio-quiet quasars that will be radio-detected, so the typical SFR of the bulk of the radio-quiet quasars could still be significantly lower than the SFR required to produce radio emission at the $L_{144}\gtrsim10^{25}$\,W\,Hz$^{-1}$ level.

To explore the contribution of star formation further, we conducted a simple Monte Carlo (MC) simulation. Our approach is similar in spirit to the MC simulations carried out by \cite{Rosario2020}, although our method is a little different. We first discard all quasars with $L_{144}>10^{26}$\,W\,Hz$^{-1}$, and, for every quasar in the remaining sub-sample, we draw a SFR from a log-normal distribution. We convert this to a rest-frame $L_{144}$ using equation~\ref{eq:sfr} and then to an observer-frame flux density, and ask whether this source would be radio-detected at that quasar's redshift, based on its predicted total flux density. The quasars with $L_{144}<10^{26}$\,W\,Hz$^{-1}$ are mostly unresolved, so their peak and total flux densities are comparable. We find the flux limit of $0.35$\,mJy quoted by \cite{Rosario2020} reproduces the radio-detection fraction fairly accurately when applied as a cut to the total flux density, so we use this value for our Monte Carlo simulations. 
We carry out the above process 100 times for each quasar, and renormalize when comparing to the observational data. We can then build a histogram for each pair of SFR distribution moments, masking bins with $<6$ counts, and fit for the two moments using a simple $\chi^2$ minimization. 

The results of this exercise are shown in Fig.~\ref{fig:SFR}. The aim here is not to produce a statistically good fit or infer parameters, rather to ascertain if star formation is a plausible origin of the radio emission in radio-quiet quasars. We found the best-fitting parameters correspond to a median SFR of $\approx 30\,M_\odot\,{\rm yr}^{-1}$, a mean SFR of $\approx 420\,M_\odot\,{\rm yr}^{-1}$, and a standard deviation in log-space of $\approx1$ dex. The best-fitting reduced $\chi^2$ was $\chi^2/{\rm d.o.f.}\approx 2$. While not formally a statistically acceptable fit, the model reproduces the shape and normalisation of the observed data histogram. As a result, it also produces a radio-detection fraction ($\approx 13$ per cent) comparable to that observed (with luminous radio sources not included). Our approach is clearly fairly crude; it does not account for a number of systematic effects, such as the SFR dependence on redshift, contributions from winds or jets, source structure/morphology, or a different, perhaps multimodal, SFR distribution. Nevertheless, the simulations still come close to reproducing the observations, suggesting that the radio emission in radio-quiet quasars can plausibly be produced by star formation, provided that median SFRs of $\sim30\,M_\odot\,{\rm yr}^{-1}$ in all quasars, and $\sim300\,M_\odot\,{\rm yr}^{-1}$ up to 1000s of $M_\odot\,{\rm yr}^{-1}$ in the radio detected sources, can be accommodated. It is also likely that multiple different mechanisms are operating across the population, so we also discuss disc winds and weak/compact jets as sources of radio emission. 

\subsubsection{Winds}
\label{sec:winds}

Disc winds can produce radio emission through free-free emission \citep{blundell_origin_2007} or synchrotron emission from shocks \citep{stocke_radio_1992,zakamska_quasar_2014,nims_observational_2015}. In the former case, super-Eddington accretion (and/or possible clumpiness) is generally required to produce high enough radio luminosities, so while free-free emission may be important in some sources, it is generally less efficient at converting kinetic power from a wind into radiation. Considering electron acceleration at shocks, \cite{nims_observational_2015} provide an estimate of the radio luminosity produced by a wind with kinetic power $L_k\sim 0.05 L_{\rm bol}$, suggesting $L_{\rm radio} \sim 10^{-5} L_{\rm bol}$. This relationship is plotted in the left-hand panel of Fig.~\figheiilum. To get this estimate, the authors assume $1$ per cent of the shock power goes into non-thermal electrons (based on what is observed in supernova remnants) and that these electrons radiate in a $3$\,mG magnetic field. The kinetic luminosity is also uncertain and the true outflow prevalence or wind covering fraction is not known. Thus, while the assumed values are reasonable guesses, there is significant room for manoeuvre in the estimated radio luminosity. The fact that the dotted red line coincides with the cloud of radio-quiet points in Fig.~\figheiilum\ should therefore be interpreted with caution; it is a demonstration that disc winds possess sufficient power to contribute to radio emission in that regime, but does not constitute actual evidence of such a contribution. In addition, it is possible that radio emission from disc winds contaminates the FIRC, which could, for example, affect the normalisation (or slope) of the empirically derived relationship between SFR and $L_{144}$. 

\subsubsection{Weak and/or Compact Jets}
\label{sec:jets}
Jets from AGN dominate the high luminosity, radio-loud AGN population, but jets also produce radio emission at lower luminosities. For example, \cite{Mingo2019} show that, even though FR II radio galaxies are preferentially found at high luminosity, both FR I and FR II jetted sources can be found right down to $L_{144} \sim 10^{22}$\,W\,Hz$^{-1}$, significantly below the traditional FR luminosity divide. The lower-luminosity FR I and FR II sources are also more compact. In addition to the FR Is and FR IIs, various classes of {\em compact} radio galaxies have also been identified. Compact steep-spectrum sources (CSS) and gigahertz peaked sources (GPS) are quite powerful at $1.4$\,GHz and are generally unresolved in all-sky radio surveys \citep{odea_what_1991,odea_compact_1998,orienti_radio_2016,odea_compact_2020}. These sources could be young or frustrated versions of larger jetted radio galaxies, and might be responsible for some of the more powerful compact, unresolved sources in our sample. In more recent years, a population of compact, lower-luminosity sources thought to be associated with weak jets have been identified \citep[e.g.][]{sadler_local_2014,baldi_pilot_2015,baldi_fr0cat_2018}, often referred to as `FR 0' radio galaxies. In addition, excess AGN activity has been invoked to explain the differences in radio properties between red and blue quasars \citep{klindt_fundamental_2019,fawcett_fundamental_2020,Rosario2020}, and \cite{jarvis_prevalence_2019} have demonstrated that many radio-quiet quasars have their radio emission dominated by the AGN rather than star formation \citep[see also][]{white_radio-quiet_2015,white_evidence_2017,herrera_ruiz_unveiling_2016,gurkan_lofar/h-atlas:_2018, Morabito2019,Smith2020}. \cite{molyneux_extreme_2019} also find a connection between ionized galaxy-scale outflows and compact radio emission, suggesting weak or young jets as the cause. However, compact AGN radio emission does not always have to be attributable to jets and could instead be related to winds or some other AGN phenomena; indeed, even the morphological distinction between wind and jet is not clear-cut in compact sources. Nevertheless, since the population of jetted radio-loud sources is likely to extend down to lower radio luminosities, we expect some fraction of the radio-quiet sources to have their emission dominated by jets with powers a few orders of magnitude lower than those in the large, luminous sources (see right-hand panel of Fig.~\figtracks). As mentioned previously, these cannot simply be young versions of the radio-loud population if the jet power is roughly constant.

\subsubsection{The multiple possible contributors to radio-quiet emission}
\label{sec:contributions}
The above arguments, and the left- and right-hand panels of Fig.~\figheiilum\ in particular, indicate that disc winds, jets and star formation are all energetically capable of producing compact, unresolved radio emission with $L_{144}\sim10^{24-26}$\,W\,Hz$^{-1}$. It is hard to distinguish further between the possible contributors, although comparison to the radio luminosity function is instructive. At low redshift and at $1.4$\,GHz, there is a transition between star-forming galaxies and radio AGN at a break luminosity around $10^{23}$\,W\,Hz$^{-1}$ \citep{mauch_radio_2007,kimball_two-component_2011,heckman_coevolution_2014}. This break luminosity probably shifts to higher luminosities in quasar hosts \citep{condon_active_2013} and at higher redshift, because the SFR is expected to increase. The lower observing frequency should also increase individual source luminosities and a higher luminosity transition between star-forming galaxies and radio-loud AGN is indeed measured by \cite{Hardcastle2019} using LOFAR data. It is therefore plausible that the apparent break in the histogram of $L_{144}$ in Fig.~\ref{fig:SFR} at around $\log_{10} (L_{144}) \approx 26-26.4$ corresponds to the transition between star formation and AGN jets dominating the radio emission. However, there are also multiple overlapping contributions from AGN jet \citep{jarvis_accretion_2004,simpson_extragalactic_2017} -- and possibly wind -- populations. We therefore do not favour one specific scenario or single origin for the radio-quiet emission, although we discuss what might drive the observed correlations in Section~\ref{sec:rqq}. Furthermore, the relative dominance of AGN and star formation might change as a function of luminosity even in the radio-quiet population. What we can be sure of is that there are {\em at least two} overlapping populations, so as to explain the clear increase in source numbers below $10^{26}$\,W\,Hz$^{-1}$ (Fig.~\ref{fig:SFR}) as well as the opposing trends shown in Fig.~\ref{fig:rdf}. This idea is consistent with the studies referenced above and can be studied in more detail with future LOFAR data releases. 

\subsection{Stochasticity and time-scales}
\label{sec:stochastic}
At a given location in \CIV\ emission line space, quasars tend to have very similar rest-frame UV properties; however, the radio properties can differ substantially, to the extent that it is possible to find sources at both sides of the radio-quiet/radio-loud and LOFAR-detected/undetected divisions at any given location in the parameter space. One way of explaining this is that the radio and UV properties trace stochastic processes operating on different time-scales. We can explore hypothesis further by considering the relevant physical time-scales in the system. 

Whatever the energy source for the radio emission, the radiation is likely to be synchrotron emission from shock-accelerated electrons. The synchrotron cooling time for electrons with a characteristic emission frequency $\nu_c$ is given by
\begin{equation}
\tau_{\rm sync} \approx 108\, \mathrm{Myr}~\left(\frac{B}{10\,\mu\mathrm{G}}\right)^{-3/2}~\left(\frac{\nu_c}{\mathrm{144\,MHz}}\right)^{-1/2},
\end{equation}
where $B$ is the magnetic field strength. The value of $B$ here is uncertain and depends on the origin of the radio emission. For large-scale radio galaxies, magnetic field strengths of $\sim 10\,\mu$G are typical \citep{croston_x-ray_2005,harwood_fr_2016}.
For these characteristic field strengths, and for the radio frequencies considered here (144\,MHz for LOFAR and 1.4\,GHz for FIRST), we might expect the synchrotron emission to trace jet activity on $\sim$10--100\,Myr time-scales. Large radio sources also take time to grow, with typical advance speeds for FR II sources of a few per cent of the speed of light \citep[e.g.][]{harwood2017} suggesting lifetimes around $100$\,Myr for a (two-sided) LLS of $600$\,kpc. These time-scales are applicable to the radio-loud sources; in the radio-quiet sources the corresponding time-scales are different and depend on the radio-quiet mechanism. Magnetic fields in star-forming galaxies range from $\mu$G levels to 100s of $\mu$G \citep{thompson_magnetic_2006}, but in quasar winds there are few observational constraints \citep[although][adopt $\sim3$\,mG]{nims_observational_2015}. Compact, galaxy-scale jets and quasar-driven winds inevitably have shorter dynamic time-scales than large-scale radio galaxies, while a typical star formation time-scale is $\sim10$\,Myr and cannot be too much longer than this for the high SFRs considered here. 

The UV emission line properties are determined by the physics of the accretion disc, the broad-line region and the putative outflow. One relevant time-scale is the viscous time-scale in a thin $\alpha$-disc \citep[e.g.][]{shakura_black_1973,frank_accretion_2002}, given by 
\begin{equation}
\tau_{\rm visc} 
\sim 5\times10^4\,{\rm yr}~
\frac{R}{5\,\rm{ld}}
\left(\frac{\alpha_{\rm visc}}{0.1} \right)^{-1}
\left(\frac{H/R}{0.1} \right)^{-1}
\left(\frac{c_s}{10\,\rm{km\,s}^{-1}} \right)^{-1}\ ,
\end{equation}
where $\alpha_{\rm visc}$ is the standard viscosity parameter, $H/R$ is the disc aspect ratio, $R$ is the radius, and $c_s$ is the sound speed, and we have chosen typical values for $\sim$100 gravitational radii in a thin disc around a $10^9\,M_\odot$ black hole. Generally, $\tau_{\rm visc}$ is much smaller than $\tau_{\rm sync}$. The thermal and dynamical times in the accretion disc are even shorter than $\tau_{\rm visc}$ \citep{frank_accretion_2002}. 

Although the hierarchy of time-scales depends on the detailed physical picture, we can expect that radio emission from large-scale lobes and jets traces longer time-scales than the UV and optical emission from the accretion disc and BLR; this difference can be expected more trivially from the sizes of, and light travel times across, the respective emission regions. Similarly, it is fairly natural for the star formation and quasar activity time-scales to be decoupled as discussed by, e.g., \cite{pitchford_extreme_2016}. A disconnect between the radio and UV properties is reasonable if the UV properties change on time-scales shorter than, say, $\tau_{\rm sync}$; for example, if quasars `flicker', or change between different accretion states. It is well known that quasars exhibit multiwavelength variability on a wide range of time-scales \citep[e.g.][]{ulrich_variability_1997,peterson_variability_2001}, and `changing-look' quasars are a particularly dramatic example of this \citep{lamassa_discovery_2015,macleod_systematic_2016,runnoe_now_2016}. Evidence for flickering or intermittent jets can be seen in a number of large-scale radio galaxies \citep[e.g.][]{konar_spectral_2006,Turner2018,maccagni_flickering_2020,shabala2020}. There are also theoretical suggestions that quasars might accrete in sporadic episodes, perhaps lasting only $10^5$\,yr \citep[e.g.][]{king_fuelling_2007,king_agn_2015}. We suspect that understanding stochasticity and intermittency (in terms of both discs and jets) is important for explaining why two given sources can look nearly identical in their UV spectra but have totally different radio properties, a statement that can be true for both radio-loud and radio-quiet sources.

\subsection[{What physics drives the trends in CIV emission space?}]{What physics drives the trends in C\,{\sevensize IV} emission space?} 
Although there are probably a number of competing factors at work, models for the radio emission must (at least) be able to explain the results shown in Fig.~\ref{fig:rdf}: why does the radio-detection fraction increase with \CIV\ blueshift, and the radio-loud fraction decrease? Additionally, at a given location in \CIV\ emission line space, what determines whether a source is radio-loud, radio detected but radio-quiet, or radio undetected? The trend in radio-loud fraction is discussed by \cite{richards_unification_2011} and driven by the prevalence of relatively large-scale, powerful AGN jets; relevant also to this discussion is the work by \cite{kratzer_mean_2015} who investigate dependence of the radio-loud fraction and mean radio-loudness not only on \CIV\ properties, but also with redshift, optical luminosity, `Eigenvector I', mass, and colour. In contrast to the radio-loud fraction, the trend in radio-detection fraction is instead driven by the mechanism(s) powering the emission in the cloud of radio-quiet sources. We discuss each of these trends in turn with reference to the relevant physics of jets, winds and star formation. 

\subsubsection{Radio-loud fraction}
\label{sec:rlf_trend}
First, we consider the decrease of the RLF with \CIV\ blueshift, first discussed by \cite{richards_unification_2011} and now confirmed at lower radio frequencies. We assume for the purposes of this discussion that the radio emission in the majority of radio-loud sources is produced by jets. If radio jets are driven by the \cite{blandford_electromagnetic_1977} mechanism, the jet power proportionality is $Q_{\rm{BZ}} \propto a_*^2 \Phi_B^2 M_{\rm BH}^2$; here $M_{\rm BH}$ is the black hole mass, $a_*$ is the dimensionless black hole spin, and $\Phi_B$ is the magnetic flux threading the event horizon.
As well as having the available spin and magnetic energy, a number of general relativistic magnetohydrodynamic simulations (GRMHD) suggest that there are other critical ingredients like the presence of an inner disc wind to collimate the flow, and a preference for certain accretion states (see reviews by \citealt{blandford_relativistic_2019,davis_magnetohydrodynamic_2020}). 

Black hole spin is interesting to consider, since higher spins lead to higher radiative efficiencies and an increase in extreme UV (EUV) flux, both because the disc extends closer to the black hole. \hl{One possibility is that quasars with lower {\ion{C}{iv}} blueshifts have higher black hole spins} that could preferentially produce jets at low blueshifts and explain the decrease of RLF with blueshift. In addition, in high spin sources the increased EUV flux could lead to a more ionized outflow and BLR. This increase in spin would increase the \HeII\ EW and decrease the line-driving force multiplier and outflow strength. Thus, spin can go someway to explaining the observed behaviours and the apparent anticorrelation between wind and jet prevalence in \CIV\ emission line space. It is worth noting that, in X-ray binaries, winds and jets appear in distinct, and generally opposite accretion states as defined by spectral hardness and X-ray luminosity (\citealt{fender_towards_2004,ponti_ubiquitous_2012}; although see also \citealt{munoz-darias_hard-state_2019,higginbottom_thermal_2020}). \cite{kording_accretion_2006} have attempted to apply similar principles to populations of AGN, but the picture is clearly more complicated, not least because of the range in time-scales involved. None the less, the \CIV\ emission line space seems to have potential as a probe of the `disc-wind-jet' connection in AGN. 

Spin is likely to be necessary, but not sufficient, for powerful jet production in AGN. General theoretical arguments for this are given by, for example, \cite{richards_unification_2011} and \cite{blandford_relativistic_2019}. The idea is also supported observationally by the fact that many radio-quiet AGN seem to be spinning rapidly, as inferred from X-ray observations \citep{reynolds_measuring_2014} and studies based on the \cite{soltan_masses_1982} argument (\citealt{elvis_most_2002,yu_observational_2002,shankar_probing_2020}; see also \citealt{broderick_is_2011}). As well as spin, the jet power also depends on the magnetic flux threading the event horizon \citep{blandford_electromagnetic_1977,tchekhovskoy_efficient_2011,davis_magnetohydrodynamic_2020}, which can either be generated in situ \citep[e.g.][]{liska_large-scale_2018} or accumulated from the surrounding interstellar medium \citep[e.g.][]{beckwith_transport_2009}. It is not clear how the magnetic flux responsible for jet launch might affect the UV emission line properties, particularly since it may be a state-dependent property of immediate black hole environment, in contrast to the black hole spin that probably changes on longer time-scales. It is also possible that the jets are primarily launched during different accretion states to the `quasar' state, in which case the disc might have little memory of the magnetic field configuration during the jet episodes \citep[e.g.][]{sikora_magnetic_2013}. Generally speaking, the dependence of jet power on magnetic flux offers a possible cause of stochastic jet behaviour as discussed in Section~\ref{sec:stochastic}, and may facilitate the hypothesis that black hole spin increases as blueshift decreases. In this framework, spin would increase the probability of producing a powerful jet and control the radio-loud trend with blueshift. Additional physics relating to the accumulation of magnetic flux and/or jet collimation would then determine whether a powerful jet could be launched for a given black hole spin \citep[see also][]{sikora_radio_2007,rusinek_diversity_2020}. These suggestions require severe qualification; there are likely to be multiple degeneracies and it is not possible to draw clear conclusions. Overall, consideration of jet launching theory (i) provides a possible framework in which black hole spin increases towards low \CIV\ blueshifts and high \HeII\ EWs, which may be testable with photoionization models; and (ii) supports the idea that stochastic behaviour is expected, which is important for explaining the radio properties of our quasars in \CIV\ emission space.  

\subsubsection{Radio-quiet quasars and the radio detection fraction}
\label{sec:rqq}
As described above, the origin of the radio emission in radio-quiet sources is not known and is likely to be attributed to at least two mechanisms. The driving factor behind the trend of increasing radio detection with \CIV\ blueshift depends on this conclusion and the relative contribution of star formation, winds, and jets.

Disc winds might also be driving the increased detection fraction with \CIV\ blueshift, even if they are not the dominant cause of radio emission in radio-quiet quasars. If so, the explanation might be fairly straightforward: since \CIV\ blueshifts are thought to be caused by a stronger outflowing component in the BLR, it is fairly natural to expect that they produce disc winds with higher kinetic powers than low blueshift sources. An increased wind power could then produce more radio emission and increase the detection fraction. Alternatively, the probability of producing a disc wind could increase with blueshift such that there are more high blueshift objects with detectable wind-related radio emission. Disc wind models for BAL quasars are capable of producing strong UV emission lines \citep{murray_accretion_1995,matthews_testing_2016,matthews_stratified_2020}, but, 
although there have been successful attempts to model \CIV\ blueshifts \citep[e.g.][]{chajet_magnetohydrodynamic_2013,Yong2017}, we are not aware of a full radiative transfer and photoionization treatment of their formation. In order to test disc wind models for radio emission, future modelling designed to constrain the kinetic power of outflows associated with blueshifts would be useful, combined with a more detailed treatment of the expected radio emission from wind-driven shocks. 

\section{Conclusions}
We have made use of the first data release of the LoTSS to investigate the low-frequency radio emission of a sample of $\simeq$10\,500 quasars in the context of their ultraviolet properties. SDSS spectrum-reconstructions from \citet{Rankine2020} allow for reliable measurements of the \CIV\,$\lambda$1549- and \HeII\,$\lambda$1640-emission lines. Our main conclusions are as follows:

\begin{itemize}
\item We have investigated the radio properties of quasars throughout the \CIV\ emission space -- blueshift versus equivalent width. 
\hl{We have found radio-detected quasars everywhere in {\ion{C}{iv}} emission space that undetected quasars can be found} (Fig.~\ref{fig:CIVdet}).

\item LOFAR's increased sensitivity relative to surveys such as FIRST enables a unique probe of the radio-quiet population (Fig.~\ref{fig:FIRST}), discovering an increasing radio-detection fraction with increasing \CIV\ blueshift (Fig.~\ref{fig:rdf}) -- which is used to infer the strength of accretion disc winds. However, the detection fraction trend with blueshift at fixed EW is not as simple as the monotonic increase observed in the whole population (Fig.~\ref{fig:RDFhex} and accompanying text). The radio-loud fraction decreases with increasing blueshift hinting at multiple sources of radio emission that correlate differently with \CIV\ blueshift.

\item \hl{Radio-loud sources can be found across the same range of {\ion{C}{iv}} emission space as radio-quiet sources, but, consistent with earlier studies}, the largest, most luminous and most radio-loud sources are found preferentially at low blueshifts and moderately high \CIV\ EWs (Fig.~\ref{fig:multiple}).

\item Luminous radio sources are also found almost everywhere in \CIV\ emission space, but, consistent with earlier studies, the largest, most luminous and most radio-loud sources are found at low blueshifts and moderately high \CIV\ EWs (Fig.~\ref{fig:multiple}). It is difficult, however, to separate different sources of radio emission and so care should be taken when performing statistical analysis on the whole population (Fig.~\ref{fig:l144_cuts}). 
Additionally, the distribution of radio-loudness as a function of \CIV\ blueshift would suggest that the radio-loud threshold should be a function of blueshift rather than a constant number.

\item Comparing FIRST-detected sources to radio-loud sources in LOFAR (Fig.~\ref{fig:FIRST}) reveals that when a reasonable radio-loudness cut is applied, the two populations have very similar distribution in \CIV\ emission space. This result suggests that the observing frequencies are tracing similar AGN phenomena and that radio-loud sources in LOFAR that are either too faint, or have too steep spectra, to be detected by FIRST are an extension of the same population.

\item Despite the trends across the parameter space, sources can look identical in the ultraviolet and have completely different radio properties. This inability to differentiate quasars with different radio properties based solely on their ultraviolet properties suggests that the radio does not know about the current state of the accretion disc, as might be expected if the processes are stochastic and the different emission traces different time-scales of activity.

\item We find the radio properties to correlate with \HeII\ EW. There is a Baldwin effect with luminosity across the radio-quiet quasars, whereas most luminous radio sources have strong \HeII\ EW (Fig.~\figheiilum, left-hand panel). Weak \HeII\ EW sources (${\rm EW}\lesssim1$\AA) are almost guaranteed to be radio-quiet with luminosities below $10^{26}$\,W\,Hz$^{-1}$. 

\item We explore the possible origin of the radio emission. We find that star formation, disc winds and weak/compact jets are all energetically capable of producing the radio-quiet quasar emission (Fig.~\ref{fig:heii_3panel}). We explore the possible role of star formation in more detail using Monte Carlo simulations, showing that a broad distribution of SFRs with median $\approx30\,M_\odot\,{\rm yr}^{-1}$ can approximately reproduce the detection fraction and distribution of radio luminosity in the radio-quiet quasars (Fig.~\ref{fig:SFR}). 
We infer SFRs in the range of 100s to 1000s of $M_\odot\,{\rm yr}^{-1}$ in the radio-detected, radio-quiet sources.
Occam's razor might lead one to invoke star formation plus jets as the simplest explanation for the radio emission, but there is plenty of room for other drivers. Multiple overlapping contributions from winds, jets, and star formation are possible (see Section~\ref{sec:contributions}). 
\end{itemize}

\noindent Our work further demonstrates the utility of low-frequency radio data for investigations regarding the origin of radio emission, particularly in radio-quiet quasars. Combining LOFAR data with the \CIV\ emission line space has allowed us to investigate the connection between winds, jets, and star formation, as well as their relationship with the AGN accretion disc and BLR. Future data releases from LoTSS covering the full northern sky will therefore be of tremendous value in addressing these questions and studying the relationship between radio emission and the rest-frame UV emission lines in more detail.

\section*{Acknowledgements}

We thank the anonymous referee for a comprehensive review. We gratefully acknowledge the use of the following software packages: Astropy \citep{astropy_collaboration_astropy_2013,astropy_collaboration_astropy_2018}, matplotlib \citep{hunter_matplotlib:_2007}, NumPy \citep{NumPy2020}, SciPy \citep{SciPy2020}. ALR acknowledges funding via the award of a Science and Technology Facilities Council (STFC) studentship. JM acknowledges funding from a Herchel Smith Fellowship at Cambridge. PCH acknowledges funding from STFC via the Institute of Astronomy, Cambridge, Consolidated Grant. MB acknowledges funding from The Royal Society via a University Research Fellowship. 
We thank Matthew Temple and Amy Kimball for helpful discussions. 

LOFAR data products were provided by the LOFAR Surveys Key Science project (LSKSP; https://lofar-surveys.org/) and were derived from observations with the International LOFAR Telescope (ILT). LOFAR (van Haarlem et al. 2013) is the Low Frequency Array designed and constructed by ASTRON. It has observing, data processing, and data storage facilities in several countries, which are owned by various parties (each with their own funding sources), and which are collectively operated by the ILT foundation under a joint scientific policy. The efforts of the LSKSP have benefited from funding from the European Research Council, NOVA, NWO, CNRS-INSU, the SURF Co-operative, the UK Science and Technology Funding Council and the Jülich Supercomputing Centre.

Funding for the Sloan Digital Sky 
Survey IV has been provided by the 
Alfred P. Sloan Foundation, the U.S. 
Department of Energy Office of 
Science, and the Participating 
Institutions. 
SDSS-IV acknowledges support and 
resources from the Center for High 
Performance Computing at the 
University of Utah. The SDSS 
website is www.sdss.org.
SDSS-IV is managed by the 
Astrophysical Research Consortium 
for the Participating Institutions 
of the SDSS Collaboration including 
the Brazilian Participation Group, 
the Carnegie Institution for Science, 
Carnegie Mellon University, Center for 
Astrophysics | Harvard \& 
Smithsonian, the Chilean Participation 
Group, the French Participation Group, 
Instituto de Astrof\'isica de 
Canarias, The Johns Hopkins 
University, Kavli Institute for the 
Physics and Mathematics of the 
Universe (IPMU) / University of 
Tokyo, the Korean Participation Group, 
Lawrence Berkeley National Laboratory, 
Leibniz Institut f\"ur Astrophysik 
Potsdam (AIP), Max-Planck-Institut 
f\"ur Astronomie (MPIA Heidelberg), 
Max-Planck-Institut f\"ur 
Astrophysik (MPA Garching), 
Max-Planck-Institut f\"ur 
Extraterrestrische Physik (MPE), 
National Astronomical Observatories of 
China, New Mexico State University, 
New York University, University of 
Notre Dame, Observat\'ario 
Nacional / MCTI, The Ohio State 
University, Pennsylvania State 
University, Shanghai 
Astronomical Observatory, United 
Kingdom Participation Group, 
Universidad Nacional Aut\'onoma 
de M\'exico, University of Arizona, 
University of Colorado Boulder, 
University of Oxford, University of 
Portsmouth, University of Utah, 
University of Virginia, University 
of Washington, University of 
Wisconsin, Vanderbilt University, 
and Yale University.


\section*{Data Availability}

The data underlying this article were accessed from the Sloan Digital Sky Survey (\url{https://www.sdss.org/dr14/}) and the LOFAR Two-metre Sky Survey (\url{https://lofar-surveys.org/releases.html}). The derived data generated in this research will be shared on reasonable request to the corresponding author.



\bibliographystyle{mnras}
\bibliography{LOFAR,james-refs} 



\appendix 

\section[Redshift evolution in C IV emission space]
{Redshift evolution in C\,{\sevensize IV} emission space}
\label{app:redshift}

\begin{figure*}
    \centering
    \includegraphics[width=\linewidth]{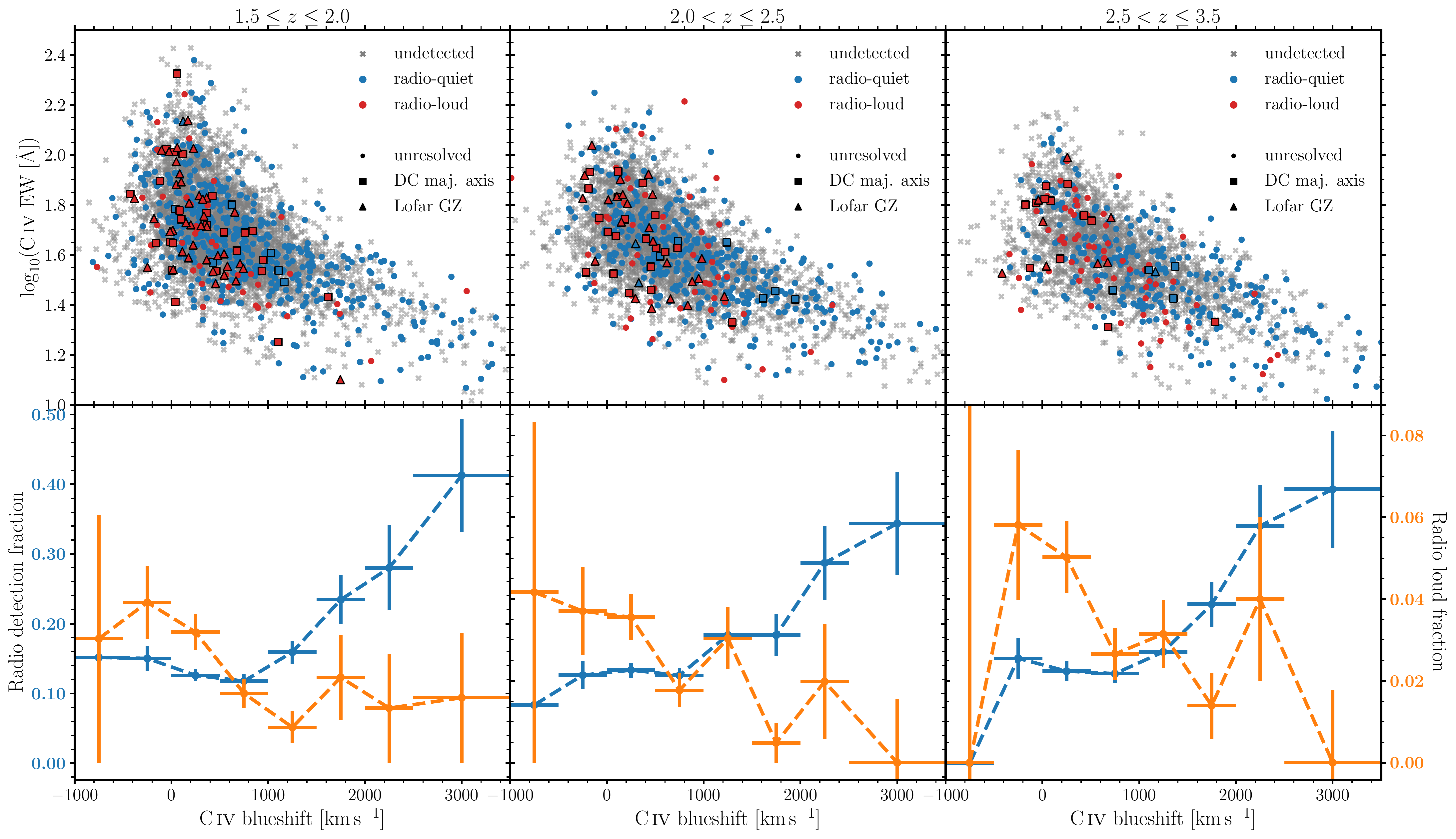}
    \caption{Top row: the \CIV\ emission space populated by quasars in the three redshift ranges. Red and blue markers indicate whether a source is radio-loud or radio-quiet (see Section~\ref{sec:CIVres}). Note that all but four of the undetected quasars (grey crosses) have radio luminosity upper limits that would classify them as radio-quiet. Marker shape denotes whether the largest linear size (see Section~\ref{sec:properties}) of the radio emission is calculated from the deconvolved major axis (square) or from the LOFAR Galaxy Zoo project (triangle), or if the emission is unresolved (circle). Bottom row: detection (left-hand axes, blue) and radio-loud (right-hand axes, orange) fractions as functions of \CIV\ blueshift for the three redshift ranges. Vertical error bars for bins empty of radio-loud quasars have been calculated assuming the presence of one radio-loud quasar. This is the case for the lowest and highest \CIV\ blueshift bins in the highest redshift subsample, and the highest \CIV\ blueshift bin in the middle redshift subsample.}
    \label{fig:z_evo}
\end{figure*}

We have split the quasar sample into three redshift bins: [1.5, 2.0], (2.0, 2.5] and (2.5, 3.5], comprising 4421, 3391 and 2313 quasars, respectively, to investigate the redshift distribution in \CIV\ emission space (top panels of Fig.~\ref{fig:z_evo}). In the lowest redshift bin, there is full coverage of the \CIV\ emission space such that the detection and radio-loud fractions as functions of \CIV\ blueshift (bottom row of Fig.~\ref{fig:z_evo}) are in closest agreement with the full sample and Fig.~\ref{fig:rdf}. As redshift increases, the highest \CIV-EW sources -- which are lower luminosity than low-EW sources -- are lost from the sample; however, the detection and radio-loud fractions remain in qualitative agreement with the full sample. Also of note and a result of LOFAR's finite resolution is the change from most radio-loud sources having resolved emission in the lowest redshift bin ($\simeq$59 per cent of radio-loud sources) to being mostly unresolved in the highest redshift bin ($\simeq$24 per cent are resolved).

\section{Effect of target selection on detection and radio-loud fractions}
\label{app:selection}

\begin{figure}
    \centering
    \includegraphics[width=\linewidth]{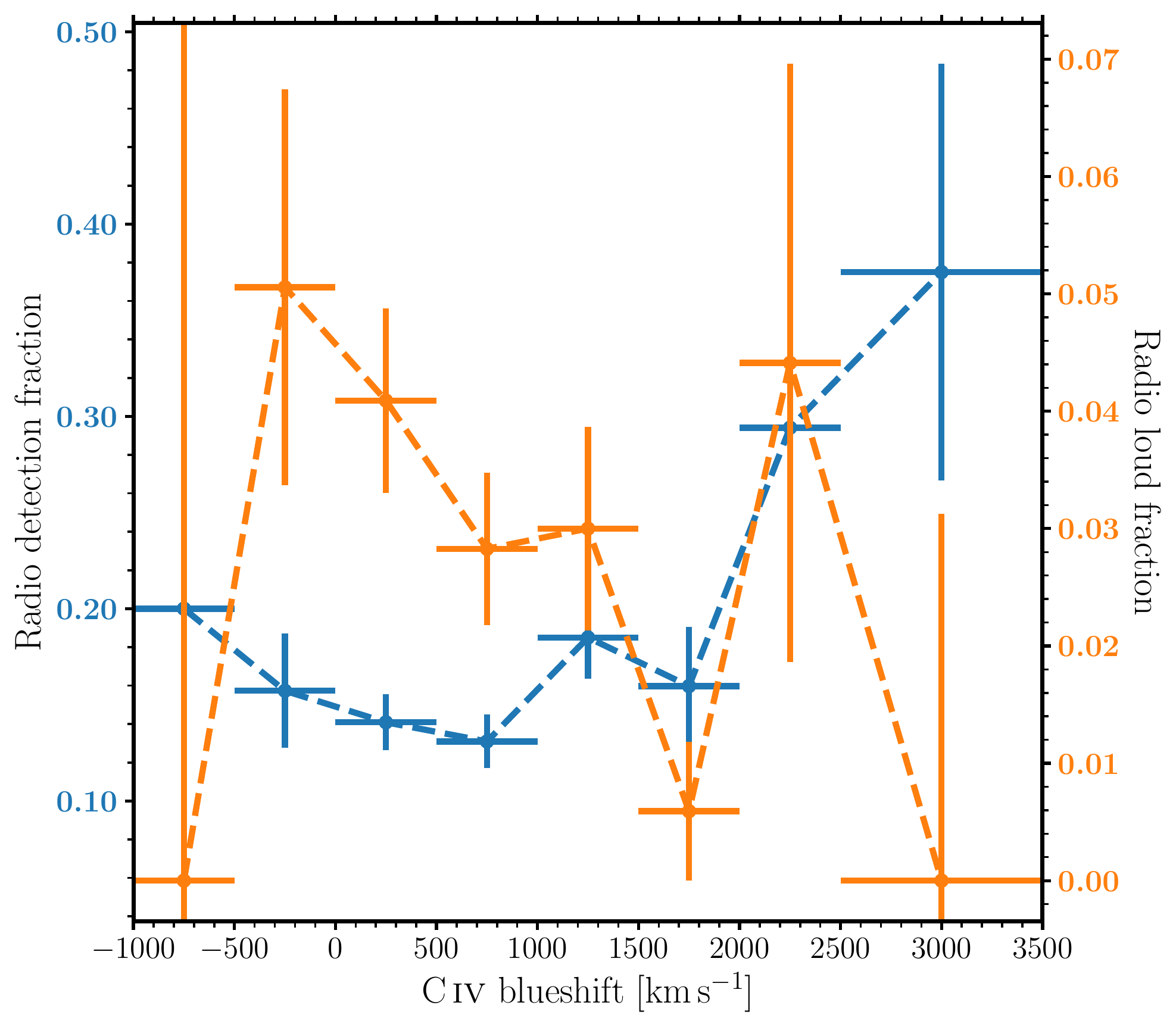}
    \caption{Same as Fig.~\ref{fig:rdf} but having limited the sample to quasars only targeted as part of the CORE BOSS sample. The lowest and highest \CIV\ blueshift bins contain zero radio-loud quasars, thus their vertical errorbars are estimated based on one radio-loud quasar populating these bins. An overall increase in detection fraction and decreasing radio-loud fraction as blueshift increases are observed.}
    \label{fig:rdfcore}
\end{figure}

We have limited the sample to the quasars targeted as part of the CORE BOSS sample, reducing the sample from 10\,163 to 2195 quasars. The 7968 quasars removed included 14 quasars targeted for their FIRST detections. Figure~\ref{fig:rdfcore} contains the radio-detection and radio-loud fractions as a function of blueshift for this limited sample and can be compared directly to Fig.~\ref{fig:rdf}. Due to the decreased sample size, the error bars are much larger than in Fig.~\ref{fig:rdf}. Limiting the sample in this way has not changed the qualitative trends illustrated in Fig.~\ref{fig:rdf}: the detection fraction increases with \CIV\ blueshift whilst the radio-loud fraction decreases with increasing blueshift.


\label{lastpage}
\end{document}